\pdfoutput=1

\documentclass[10pt,sigconf,letterpaper]{acmart}

\usepackage{graphicx}
\usepackage{enumitem}
\usepackage[font=small,skip=5pt]{caption}
\usepackage{tikz}
\usetikzlibrary{arrows.meta,positioning,calc}
\usepackage{comment}
\usepackage{xspace}
\usepackage{csquotes}
\usepackage{booktabs}
\usepackage[font=small,skip=5pt]{subcaption}
\usepackage{algorithm}
\usepackage{algpseudocode}
\usepackage{tabularx}
\usepackage{multirow}
\usepackage{soul}
\usepackage{wrapfig}
\usepackage{xcolor}
\usepackage{colortbl}
\usepackage{todonotes}

\usepackage{tcolorbox}

\usepackage{cleveref}

\DeclareCaptionLabelFormat{custom}{#2)}
\newcommand{\momq}{\texttt{MoMQ}\xspace}
\newcommand{\moqt}{\texttt{MoQT}\xspace}
\newcommand{\mpquic}{\texttt{MPQUIC}\xspace}
\newcommand{\iframe}{I-frame\xspace}
\newcommand{\iframes}{I-frames\xspace}
\newcommand{\pframe}{P-frame\xspace}
\newcommand{\pframes}{P-frames\xspace}
\soulregister{\momq}{0}
\soulregister{\moqt}{0}
\soulregister{\mpquic}{0}
\soulregister{\iframe}{0}
\soulregister{\iframes}{0}
\soulregister{\pframe}{0}
\soulregister{\pframes}{0}

\newcommand{\rulemark}[1]{{\setlength{\fboxsep}{1.2pt}\setlength{\fboxrule}{0.5pt}\fbox{\footnotesize\textbf{#1}}}\xspace}
\newcommand{\rone}{\rulemark{R1}}
\newcommand{\rtwo}{\rulemark{R2}}
\newcommand{\rthree}{\rulemark{R3}}
\newcommand{\rfour}{\rulemark{R4}}
\soulregister{\rone}{0}
\soulregister{\rtwo}{0}
\soulregister{\rthree}{0}
\soulregister{\rfour}{0}

\usepackage{pifont}
\definecolor{chalone}{HTML}{6A00FF}
\definecolor{chaltwo}{HTML}{60A917}
\newcommand{\chone}{\textcolor{chalone}{\ding{182}}\xspace}
\newcommand{\chtwo}{\textcolor{chaltwo}{\ding{183}}\xspace}

\newcommand{\afblock}[1]{\smallskip\noindent\textbf{#1}\hspace{0.4em}\ignorespaces}

\newtcolorbox{takeaways}{
  colback=black!4,
  colframe=black!45,
  boxrule=0.5pt,
  arc=2pt,
  left=5pt, right=5pt, top=4pt, bottom=4pt,
  boxsep=0pt,
  fontupper=\small,
  before upper={\textit{Takeaway.}\hspace{0.5em}\ignorespaces}
}

\newcolumntype{Y}{>{\centering\arraybackslash}X}
\newcolumntype{L}{>{\raggedright\arraybackslash}X}

\newenvironment{tightlist}{
\begin{list}{$\bullet$}{
    \setlength{\topsep}{.5em}
    \setlength{\partopsep}{0in}
    \setlength{\parskip}{0in}
    \setlength{\itemsep}{0in}
    \setlength{\parsep}{0in}
    \setlength{\leftmargin}{.5em}
    \setlength{\rightmargin}{0in}
    \setlength{\itemindent}{0in}
}}
{\end{list}}


\crefformat{section}{\S#2#1#3}
\crefformat{subsection}{\S#2#1#3}
\crefformat{subsubsection}{\S#2#1#3}
\crefrangeformat{section}{\S\S#3#1#4 to~#5#2#6}
\crefmultiformat{section}{\S\S#2#1#3}{ and~#2#1#3}{, #2#1#3}{ and~#2#1#3}

\setcopyright{none}
\renewcommand\footnotetextcopyrightpermission[1]{}

\begin{document}

\title{Media-over-Multipath-QUIC for Realtime Video Applications}

\author{Tanya Shreedhar}
\authornote{Both authors contributed equally to this research.}
\affiliation{%
  \institution{Delft University of Technology}
  \country{The Netherlands}}

\author{Zuji Zhou}
\authornotemark[1]
\affiliation{%
  \institution{Delft University of Technology}
  \country{The Netherlands}}

\author{Nitinder Mohan}
\affiliation{%
  \institution{Delft University of Technology}
  \country{The Netherlands}}

\author{Fernando Kuipers}
\affiliation{%
  \institution{Delft University of Technology}
  \country{The Netherlands}}

\renewcommand{\shortauthors}{Shreedhar and Zhou et al.}

\begin{abstract}
Multipath transports place a client's WiFi, cellular, and satellite networks under one connection, yet real-time video gains little from them.
The scheduler that assigns packets to paths sees only bytes, so it cannot tell a keyframe that anchors a second of video from an enhancement frame whose loss costs one image.
We show that the limiting factor is not a shortage of path diversity but the absence of a channel through which the application can name what the transport cannot see.
Media over QUIC Transport (\moqt), already deployed on production CDNs, carries media as named Objects, and relays forward each Object's metadata without interpreting it.
The knowledge the scheduler lacks therefore already flows through the subscriber's relay.
We present \momq, a \moqt extension that turns this metadata into path decisions.
Applications and relay operators install declarative rules at the edge relay that match Object metadata and express delivery preferences against labeled paths.
The relay evaluates the rules mechanically, so it acts on video semantics while containing no video logic.
On a live testbed spanning Starlink and WiFi, four rules cut P99.9 frame completion time from 384.7~ms under the best transport-only scheduler to 114.1~ms and reduce the required playback buffer by 61.5\%.
\momq is the only configuration that meets the 150~ms interactive latency target.
Across relays in two countries and subscribers on two continents, the same rules apply unchanged and retain their advantage wherever the paths remain disjoint.
\end{abstract}

\maketitle

\section{Introduction}
\label{sec:introduction}

Real-time video spans live streaming, conferencing, and cloud gaming, and its most interactive uses require a frame to travel from capture to display within roughly 150~ms~\cite{itu-g114}.
A single access network leaves little slack inside this budget, because any congestion episode or wireless disruption translates directly into late frames.
Each such frame misses its playout deadline, and the receiver stalls playback or discards it.
Client endpoints increasingly hold several access networks at once.
Laptops and phones combine WiFi with cellular, and LEO satellite service adds another whose failures are largely independent of terrestrial ones~\cite{ramirez-multiconnectivity}.
These networks also differ in cost.
Starlink, for instance, charges a flat rate while a terrestrial one is metered.

Past efforts to use these networks together add multipath capability inside the transport layer, so that several attachments work as one end-to-end connection.
MPTCP does so for TCP and is widely deployed in production~\cite{raiciu-mptcp, rfc8041, mptcpScan-ifip}.
\mpquic brings the same capability to QUIC, and is currently in standardization discussions~\cite{mpquic-draft, deconinck-mpquic}.
Both extensions rely on a scheduler that assigns each packet to a path.
It treats the application-layer packets as uniform bytes and decides from transport metrics alone, namely round-trip time, loss, and congestion-window state.
MPTCP forces this blindness, since it lives in the kernel behind a socket interface that offers applications no way to express preferences about path management or data scheduling~\cite{rfc6897}.
As a result, the schedulers explored so far largely port the logic of MPTCP into \mpquic, from round-robin and lowest-RTT heuristics to blocking-aware refinements~\cite{blest, ecf, paasch-schedulers}.
\mpquic lifts this placement barrier, since a QUIC stack is also implemented at the application layer.
However, the scheduling abstraction remains largely unchanged.
Data still enters the transport as ordered streams of opaque bytes, and no interface lets an application state what those bytes need~\cite{mpquic-draft}.

Real-time video workloads also increasingly adopt scalable video coding (SVC)~\cite{svc-overview}.
Its intra-coded IDR (Instantaneous Decoder Refresh) frames, or \emph{I-frames}, carry a complete picture, and \pframes encode only differences from earlier reference frames (see \cref{fig:momq_challenges}).
An \iframe is roughly $23\times$ larger, enough to occupy a path for several congestion-window rounds~\cite{wiegand-avc}.
A \pframe is small but undecodable until its references arrive.
Delivering such a stream over multiple heterogeneous paths is a scheduling problem, since a large \iframe should avoid a path about to be disrupted (or delayed), dependent frames should stay together, and a metered path should only be used when required.
Existing \mpquic schedulers cannot honor this, since no transport signal marks frame boundaries, criticality, or dependencies.

Media over QUIC Transport (\moqt), a protocol under active standardization at the IETF~\cite{moq-transport}, offers a place to close this gap.
\moqt carries real-time media for conferencing and streaming alike as named Objects through a tree of relays.
One publisher connection serves many subscribers, and each relay caches, prioritizes, and forwards Objects close to the receivers.
Cloudflare already runs a \moqt relay on every server of its global CDN, and an industry consortium including Akamai, Cisco, and Google is building production relays~\cite{cloudflare-moq, cloudflare-moq-relays, openmoq}.
This tree's last hop, from edge relay to subscriber, crosses the access networks above.
On that hop the relay is the sender, so the scheduler that chooses among those networks must run there.
Prior application-aware schedulers build the logic into the sender's own transport~\cite{xlink}.
A relay serves several applications simultaneously, so it cannot embed the logic of any one of them.
Each Object, however, carries application metadata that the relay forwards without interpreting, yet nothing in \moqt relates an Object to a path.

We present \momq, a \emph{compatibility} layer between \moqt and \mpquic that enables multipath real-time video delivery over subscriber's last-mile access networks.
The subscriber installs declarative rules on its edge relay that map Objects onto suitable paths, and the same interface opens \moqt-level scheduling optimizations to CDN operators.
To the best of our knowledge, \momq is the first multipath extension to \moqt. 
Specifically, we make the following contributions.
\begin{tightlist}
    \item We design \momq, a multipath extension that operates as a layer between \moqt and \mpquic (\cref{sec:design}). \mpquic keeps path management, congestion control, and packet scheduling, and \momq passes it the application metadata that those mechanisms cannot see. Rules refer to paths through descriptive labels rather than identifiers, so a policy carries across access technologies and new scenarios need new labels and rules rather than protocol changes. The extension adds five wire-level elements and falls back to standard \moqt when a session does not negotiate it.
    \item We define the scheduling rules and environment-specific enhancements that adapt \momq to widely used access networks (\cref{sec:casestudy}). A four-rule policy derived from live measurements addresses the scheduling failures of SVC conferencing over satellite and terrestrial paths, where the terrestrial path models a metered access such as cellular. A reconfiguration-avoidance enhancement exploits the predictable disruption schedule of LEO constellations.
    \item We implement\footnote{To be released upon acceptance.} \momq and evaluate it on a live testbed with subscriber nodes in Europe and North America, each holding a Starlink terminal alongside a terrestrial link (\cref{sec:evaluation}). The four-rule policy reduces P99.9 frame completion time from 384.7~ms under the best transport-only scheduler to 114.1~ms, the only delivery tail inside the 150~ms budget. It uses the metered path for 11.3\% of traffic, against 90--100\% for the latency-oriented schedulers. The gains persist at $2.5\times$ the relay distance and vanish when the two paths converge onto a shared backbone.
\end{tightlist}

\section{Background and Related Work}
\label{sec:background}
\label{sec:motivation}

\subsection{Background}
\label{subsec:background}

\label{subsec:bg_moqt}

\afblock{Media over QUIC Transport} (\moqt) is in active standardization at the IETF.
It organizes media as a hierarchy of Track, Group, Subgroup, and Object, whose levels map in layered video to a resolution layer, a Group of Pictures, a temporal layer, and an individual frame~\cite{moq-transport}.
Each Object carries metadata alongside its immutable payload, and applications attach their own metadata through code point ranges reserved for application use.
Relays must forward this metadata unchanged without attempting to interpret its meaning.
A companion draft makes this boundary cryptographic by authenticating metadata end to end, so a relay can read it but not alter it~\cite{moq-secure-objects}.
A two-level priority scheme orders delivery under congestion, with subscriber priority expressing receiver preferences across Tracks and publisher priority expressing the inherent importance of an Object or Subgroup.

Earlier delivery architectures offer two kinds of intermediary.
A Selective Forwarding Unit (SFU) selects and forwards at frame granularity by parsing the media~\cite{gso-simulcast, scallop, rfc9626}, while a CDN cache scales because it stores opaque bytes and sees nothing finer than a segment~\cite{nygren-akamai, bentaleb-abr}.
A \moqt relay observes per-Object structure as an SFU does while remaining as independent of the media format as a cache, and the base protocol guarantees both properties.
\moqt also lets a subscriber install property filters that the relay evaluates against Objects as it forwards them, and multiple filters combine into compound conditions~\cite{moq-transport}.
Every filter action, however, resolves to forwarding an Object or withholding it, and the specification does not model a relay with more than one downstream path.
\momq adopts this declarative matching and binds it to a new action, path selection (\cref{sec:design}).

\label{subsec:bg_mpquic}

\afblock{Multipath QUIC} (\mpquic)~\cite{mpquic-draft, deconinck-mpquic} runs one QUIC connection over several network paths in parallel, with independent RTT estimates, congestion state, and loss detection per path.
Existing schedulers pick a path by transport heuristics, distributing packets cyclically (Round-Robin), by lowest observed round-trip time (MinRTT), by predicted head-of-line blocking (BLEST~\cite{blest}), or redundantly on every path, with further variants refining the same signals~\cite{ecf, paasch-schedulers, shreedhar2018qaware}.
The specification leaves scheduling to the implementation, guided by preferences the application is assumed to hold, yet it defines no interface through which those preferences could be expressed~\cite{mpquic-draft}.
Running \moqt over out-of-box \mpquic, the scheduler remains blind to the Object boundaries and provides application-agnostic delivery.

\label{subsec:four_problems}
\label{subsec:bg_svc}
\label{subsec:mpquic_fails}

\afblock{Scalable Video Coding} (SVC)~\cite{svc-overview} encodes a video once into a base layer and enhancement layers that refine it, so a receiver decodes as many layers as its network sustains.
Deployed conferencing platforms including Google Meet and Jitsi Meet already run SVC in production~\cite{hancke-av1-meet,jitsi-svc}, which sharpens the semantic gap between \moqt and \mpquic.
\Cref{fig:momq_challenges} shows the temporal layering of the three-layer stream we study, in which the base layer alone plays at 13~fps and the full stack at 50~fps.
The encoder partitions the stream into Groups of Pictures (GOPs), self-contained runs of frames that decode without reference to anything outside the group.
Each GOP opens with an \emph{I-frame} (IDR), which carries a complete image.
Every other frame is a \pframe that encodes only its difference from an earlier reference frame, so it is undecodable until that reference arrives (see \cref{fig:momq_challenges}).

\begin{figure}[t]
\centering
\includegraphics[width=\columnwidth]{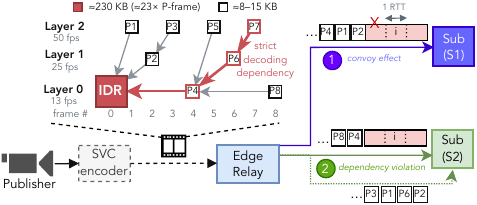}
\caption{SVC delivery through an \moqt edge relay. Left, the temporal layering of one GOP, with arrows highlighting frame references. (\chone)~\pframes queued behind a multi-round IDR transmission on a single path. (\chtwo)~Dependent frames split across two paths and arriving out-of-decode-order.}
\label{fig:momq_challenges}
\end{figure}

This structure leaves the frames of one stream deeply unequal (\cref{app:encoding} gives the per-type breakdown).
At roughly $23\times$ the size of a \pframe~\cite{wiegand-avc}, an \iframe exceeds the congestion window, so its transmission spans several congestion-window rounds while a \pframe completes within one.
Because every frame depends on its \iframe directly or transitively, losing an \iframe invalidates its entire GOP, a full second of video, while losing a top-layer \pframe costs a single image.
Enhancement frames can therefore tolerate risk that base-layer frames and \iframes cannot.
A frame that arrives before its reference is undecodable until the reference follows, so frames that depend on one another must also arrive close together.
None of these distinctions survives below the Object layer.
On the wire an \iframe and a \pframe are indistinguishable, so any differentiated treatment requires additional context at the transport.

In the scenario of \cref{fig:momq_challenges}, an edge relay forwards the stream to subscribers whose network attachments differ, with S1 reachable over one path and S2 over two.
For S1, the frame inequality first surfaces as the \emph{convoy effect} (\chone in \cref{fig:momq_challenges}).
An \iframe that exceeds the congestion window drains over several transmission rounds, and every \pframe produced while it drains is queued behind it.
Each queued \pframe inherits several RTTs of delay, and the convoy recurs at the head of every GOP.
The convoy deepens when the path drops packets, whether through wireless channel losses~\cite{narayanan-5g-variegated} or through the scheduled reconfigurations of LEO satellite networks~\cite{mohan-starlink, dissectstarlink2026sigcomm}.
Every retransmission extends the \iframe's occupancy of the path.
A reconfiguration met mid-transmission, the \emph{reconfiguration collision}, can raise the \iframe's completion time to several times the interactive budget.
\moqt's static priorities cannot express the remedy, because the correct ordering is conditional.
\pframes should overtake only while an \iframe occupies the path, and only within a budget that does not delay the \iframe itself.

These failures invite a second path.
For S2, consider the pairing of a LEO satellite link with a terrestrial link, since the two differ in delay, stability, per-byte cost, and failure behavior~\cite{ramirez-multiconnectivity,mohan-starlink} (our measurement setup in \cref{subsec:eval_setup}).
The satellite path offers wide coverage and high capacity at a flat rate, subject to the reconfigurations above.
The terrestrial path offers lower and steadier delay over a narrower link that is usually metered.
However, splitting the stream across the two paths adds two challenges.
The first is the \emph{dependency violation} (\chtwo in \cref{fig:momq_challenges}).
When a multipath scheduler splits dependent frames across paths with asymmetric RTTs, the receiver must buffer every early-arriving frame for at least the RTT difference until its reference arrives on the slower path.
Recall that the multipath scheduler cannot see these SVC dependencies.
The second is \emph{backup overuse}, because access cost is equally invisible to the transport.
Latency-optimizing schedulers concentrate traffic on the lower-RTT metered backup, and the user pays per-byte charges for minor latency gains.

All challenges described above share the same root cause.
Every property that separates one frame from another, its size class, its dependents, its tolerance for delay or loss, exists only above the layer where the path is chosen.
As we show in \cref{subsec:eval_transport}, transport heuristics do not compensate and
no existing \mpquic scheduler meets the interactive latency target.
We argue that last-mile path diversity is not the bottleneck, and
the missing ingredient is application context rather than a better transport heuristic.

\subsection{Related Work}
\label{sec:related}
\label{subsec:related}

Multipath transport for real-time media has a long lineage, and in all of it the application-aware decision runs at a sender that owns the traffic.
MPRTP schedules RTP media across paths at the sender~\cite{mprtp}, and XLINK drives a production \mpquic scheduler with video QoE feedback across three million short-video plays~\cite{xlink}.
AUGUR schedules frame retransmission and path switching for cloud gaming from per-user probability models at the service's own edge servers~\cite{augur}, and Converge builds a video-aware multipath scheduler for WebRTC with path-specific packet protection~\cite{converge}.
QCON reaches multi-connectivity from inside the 5G radio access network instead, and steers streams across base stations below the end-to-end transport~\cite{qcon}.
COMPACT splits foreground and background video tiles across cellular paths from a sender-side classifier~\cite{compact}, and ALCS compensates latency asymmetry for MPTCP on the same satellite and terrestrial pairing we study~\cite{alcs}.
Transport-only schedulers such as BLEST and ECF avoid application coupling by using transport signals alone, and remain blind to what the bytes mean~\cite{blest, ecf}.
On the LEO side, prior work shows that congestion control misreads satellite mobility as congestion~\cite{lai-leo-mobility}, LeoCC makes a single satellite path robust to that dynamic~\cite{leocc}, and handover-aware pacing resizes a WebRTC sender queue ahead of predicted disruptions~\cite{gottipati-pacing}.
Casparsen et al.\ predict degraded LEO windows and suggest steering traffic between interfaces during them, without a mechanism that could act on the suggestion~\cite{casparsen-leo-latency}.

A second line gives applications direct influence over multipath scheduling.
ProgMP lets an application install its own MPTCP scheduling program in the sender's kernel~\cite{progmp}, and DTP attaches size, priority, and deadline metadata to blocks that its endpoint transport schedules against~\cite{dtp}.
Stream-aware \mpquic schedulers map streams to paths by priority~\cite{rabitsch-stream, pstream}, MP-DASH states path preferences bound to DASH segment deadlines~\cite{mp-dash}, and Kozuka and Okabe bind streams to policy-defined path groups~\cite{kozuka-policy}.
Cech et al.\ rethink the \mpquic scheduling interface itself, with the application pacing transmission through tokens~\cite{cech-tokens}, and an Internet-Draft maps the modalities of a multimodal stream to paths at the endpoint transport~\cite{camp-multipath}.
On a single path, Salsify couples codec and transport into one control loop~\cite{salsify}, and Syntra synthesizes cross-layer video controllers from a declarative objective~\cite{syntra}.
The interfaces range from imperative programs to declarative objectives, but every system in this line assumes the scheduler runs where the application does.
\momq moves this decision to the CDN-operated edge relay.
The relay serves many applications at once and cannot be trusted to run their code, so preferences reach it as declarative rules over per-Object metadata.

Within MoQ, recent work treats the relay as a general object substrate, both by applying it beyond media~\cite{moq-dns} and by proposing selective Object dropping at the relay~\cite{moq-filtering}.
The closest MoQ work to ours reaches multiple networks only through sequential failover, where a subscriber holds connections to several relays and switches on transmission error~\cite{nemeth-failover}.
Blind duplication across paths needs no protocol support, and it appears in our evaluation as the Redundant baseline, at twice the bandwidth and with no per-Object decision.
The \moqt specification itself never mentions multiple paths and treats redundancy only as relay and publisher duplication with sequential switchover~\cite{moq-transport}.
We are aware of no Internet-Draft that proposes multipath for \moqt.
To our knowledge, \momq is the first system in which a \moqt relay selects among concurrent paths per Object.

\section{\momq{}: Media over Multipath QUIC}
\label{sec:overview}
\label{sec:design}

\momq is a backward-compatible extension to \moqt that lets the relay schedule Objects across the paths of an \mpquic connection to the subscriber.
Every Object passes through the relay that makes the forwarding decision, and the application can attach descriptive metadata to each Object, for instance a field that distinguishes an \iframe from an enhancement-layer \pframe.
The metadata sits as key-value pairs in a block between the unchanged Object header and the payload, which the relay never reads (\cref{fig:momq_object_format} in \cref{app:format}).
\moqt, however, lacks a channel through which the application can state what should happen to an Object carrying a given metadata value, and \momq adds this channel.

The subscriber installs rules on its edge relay that match generic Object metadata and express delivery preferences against labeled paths rather than path identifiers (\cref{fig:momq_stack}).
For example, a rule can keep every Object that depends on another Object on the same path as its dependency, or prefer paths of a given cost class (\cref{subsec:svc_rules} derives four such rules for SVC delivery over LEO and terrestrial paths).
The relay evaluates the rules mechanically as Objects pass through it and merges the matching actions into a per-Object directive that its \mpquic scheduler takes as advice.
It never interprets what a metadata value means, so it acts on video semantics while containing no video logic.
The rule interface follows the match-action abstraction of programmable forwarding~\cite{mckeown-openflow, p4}, applied one layer up.
The endpoints keep the semantics and the transport implements only the mechanism~\cite{saltzer-e2e, clark-alf}.

\begin{figure}[t]
\centering
\includegraphics[width=\columnwidth]{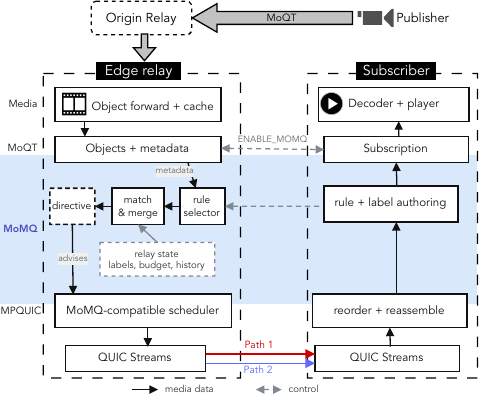}
\caption{\momq at the last hop of an end-to-end \moqt delivery. The subscriber installs rules and path-label annotations over the \moqt session. The edge relay matches each Object's metadata against them, merges the matching actions into a per-Object directive, and hands it to the \mpquic scheduler as advice rather than a command. Upstream hops are unmodified \moqt.}
\label{fig:momq_stack}
\end{figure}

\label{subsec:architecture}
\afblock{Last-hop architecture.}
\momq operates only at the last hop of the relay tree, where the edge relay and the subscriber may hold an \mpquic connection across the subscriber's access networks\footnote{\Cref{sec:discussion} discusses extending \momq to the publisher's first hop.}.
Upstream hops carry standard \moqt unchanged, since they traverse backbone networks whose path diversity the infrastructure manages.
The subscriber, by contrast, knows the cost and capacity of each of its own access networks.
Media flows downstream, so the relay is the sending endpoint of the multipath connection and the only party that holds the Objects and chooses their paths.
Only the edge relay implements the extension, and because each subscriber and relay pair runs an independent \momq session, one relay serves \momq-enabled and legacy subscribers at once.
This placement assumes one relay reachable over all of the subscriber's access networks, and an edge relay in a well-connected datacenter meets this assumption.
The alternative of connecting to a separate relay inside each access network needs no protocol change, but it forfeits the per-Object path choice and the shared cache of a single multipath sender.

\afblock{Design challenges.}
\label{subsec:challenges}
The first challenge is \emph{expressiveness against agnosticism}.
The interface must encode scheduling policies such as keeping dependent SVC frames together and steering \iframes away from a disruption window.
It must simultaneously stay blind enough that the relay never parses a codec, never tracks a dependency graph, and never needs a software update when a new format appears.
The second is that rules must \emph{outlive the topology}.
A subscriber that moves from WiFi and cellular to WiFi and satellite should not have to rewrite its policy, yet any rule that names a path directly is invalidated the moment paths are added, removed, or renumbered.
The interface therefore cannot expose path identifiers and needs a level of indirection between the application's stated preference and the path that satisfies it.

The third is that the relay must \emph{retain authority}.
It observes live path conditions that the subscriber cannot see, serves many subscribers at once, and is subject to operator policy.
An application that could demand a specific path could starve other subscribers or force traffic onto a path the operator has reserved, so application preferences must be advisory.
The fourth is that evaluation must be \emph{cheap and bounded}, because it runs once per Object on the forwarding path.
An interface permitting arbitrary predicates invites unpredictable latency and denial of service through deliberately expensive matches, so every relay must support the policy language.

\afblock{Rules over opaque metadata.}
\label{subsec:rules}
A rule contains match entries that select Objects by metadata, action entries that express delivery preferences, and a priority that orders it against the other installed rules.
The relay's role is confined to matching conditions and executing actions, so a relay that appears to understand SVC is one configured with rules that match SVC frame types (the first design challenge).
The match language has two operators (\cref{tab:match_operators} in \cref{app:format}), where \texttt{EQUALS} compares a metadata value byte for byte against a stated value and \texttt{EXISTS} tests for the presence of a key.
Match entries within a rule combine with AND, disjunction is expressed by installing several rules, and a rule with no match entries is a catch-all.
A match entry ranges over the Object's metadata joined with annotations that the relay's own components may inject at evaluation time, for instance a flag that a link's disruption window is currently open.
A rule can therefore condition on state the relay observes locally through the same two operators (\cref{subsec:svc_rules} injects two such annotations for LEO satellite paths).

Both operators reduce to byte comparisons, so rule evaluation runs at a cost bounded by the number of installed rules and a match cannot be made deliberately expensive (recall fourth design challenge).
\texttt{EQUALS} covers categorical distinctions such as frame type and layer index, \texttt{EXISTS} covers presence-based policies such as acting on every Object that declares a dependency, and more complex predicates are expressed through rule composition.
The complete policy of our case study uses only these two operators (\cref{subsec:svc_rules}).
The language cannot express negation or a numeric range.
A policy that needs an else branch obtains it from priority order, since an Object that fails every specific match falls through to the catch-all.
Evaluation cost still grows with the number of installed rules, so the design also recommends per-session caps on rule count, match entries, key and value lengths, and installation rate (\cref{app:limits} gives recommended values).
The relay enforces a cap by rejecting the offending operation.

\texttt{PRIORITY} governs the order in which queued Objects reach the wire under contention (\cref{tab:action_types} in \cref{app:format} lists all four action types).
\texttt{BALANCING} constrains whether one Object's packets may split across paths for throughput.
Its default keeps each Object on a single path, since cross-path reordering is the primary source of the application-level latency degradation observed in \cref{subsec:mpquic_fails}.
\texttt{PATH\_PREFERENCE} biases selection toward paths whose labels contain a specified key-value pair, and falls back to default selection when no healthy path matches.
A rule preferring \texttt{cost\_class=free} therefore does not block delivery when the free path is temporarily unavailable.
\texttt{PATH\_AFFINITY} names a metadata key whose value holds the \texttt{object\_id} of another Object.
The relay looks that identifier up in a bounded history of recent assignments (one byte comparison) and sends this Object on the path it recorded for the referenced one.
When multiple rules match an Object, their actions merge per dimension, taking the maximum priority, a single path if any rule requests one, and the first matching affinity.
Path preferences merge as a union ordered by rule priority.
When two rules prefer different values of the same label key, the higher-priority rule prevails, so a specific rule overrides a catch-all default.
An action entry carries its own length and a relay skips an action type it does not implement.
This design choice leaves the door open for future extensions as new delivery preferences are expressed as new action types rather than protocol revision.

\afblock{Metadata and path labels.}
\label{subsec:metadata}
The publisher attaches metadata to an Object as key-value pairs, which travel with it to every relay that forwards it.
The pairs form a block between the unchanged Object header and the payload (see \cref{fig:momq_object_format} in \cref{app:format} for the \momq object layout.
\momq defines well-known keys such as \texttt{frame\_type} and \texttt{temporal\_layer}, together with the addressing key \texttt{object\_id} that \texttt{PATH\_AFFINITY} resolves against.
Applications may define their own keys, which relays match without interpretation.
The publisher that attaches a key and the subscriber that matches on it are two ends of the same application, so they share the vocabulary without relay involvement.
Application keys sit in the ranges \moqt reserves for application use (\cref{sec:background}), so the relay's blindness is guaranteed by the base protocol, not asserted by our design extension.

Path labels describe path properties and come from both ends of the connection.
The relay labels paths from operator configuration or transport observation, for instance \texttt{link\_type=satellite}, and the subscriber annotates them with knowledge only it holds, for instance \texttt{cost\_class=metered}.
The relay merges both sets per path, with the subscriber's value taking precedence where the two assign the same key, so a rule can act on information that neither side holds alone.
The override is confined to descriptive keys such as \texttt{cost\_class}, so a subscriber label can never alter how the relay configures its transport (see \cref{subsec:impl_directive}).
Because rules reference labels rather than path identifiers, they remain valid when paths are added, removed, or renumbered (the second design challenge).
A rule preferring paths where \texttt{cost\_class=free} applies to whichever path carries that label at the moment, and a switch to a metered satellite plan becomes a single relabel rather than a rule reinstallation.
The indirection also hides the relay's topology from the subscriber and leaves the relay free to translate a preference into a path decision while respecting subscriber's preferences.

\afblock{Scheduling pipeline.}
\label{subsec:pipeline}
For each outgoing Object the relay executes three stages, and each stage reads distinct state.
\emph{Rule evaluation} walks the installed rules in priority order and collects the actions of every rule the Object matches, consulting only the session's rule store and the relay's current annotations.
\emph{Directive merge} resolves those actions per dimension into one scheduling directive, and consults nothing else.
\emph{Path selection} hands the directive to the \mpquic scheduler, which weighs it against path labels and live transport metrics. 
Within one directive an affinity \emph{binds} while a preference \emph{biases}, so a resolvable affinity decides the path before any preference is weighed.
The directive crosses the layering boundary as scheduler-readable context is attached to the Object's stream (see \cref{subsec:impl_directive}).

If no rule matches, the Object carries a default directive, \texttt{BALANCING}, that balances across all paths.
A directive that cannot be satisfied under current conditions yields to the default scheduler, so \momq never blocks delivery.
Every directive is \emph{advice}, so the relay keeps final authority over path selection (the third design challenge).
The relay honors a directive according to local policy and resource availability.
Note that differentiated services applies the same division, where per-hop behaviors are advisory and enforcement is domain-local~\cite{blake-diffserv}.
Advisory semantics let relays with different capabilities interoperate, and they accommodate conflicts the application cannot foresee, such as two subscribers requesting the highest priority when only one path is healthy.

The pipeline's two inputs, Object metadata and path labels, are also its extension points.
The relay may refresh labels continuously from any local source of path knowledge, such as a model of a link's disruption schedule.
A rule that references such a label converts that knowledge into scheduling behavior while the rule language, the wire format, and the evaluation logic remain unchanged.
A specialization for a link technology is therefore an add-on that any operator or application can deploy, namely a label provider that annotates paths or the match context, plus rules that reference the annotation.
\momq itself contains no environment-specific logic.
We build one such add-on for LEO satellite paths in \cref{subsec:svc_rules} and evaluate it in \cref{sec:evaluation} against state-of-the-art alternatives.

\label{subsec:protocol_surface}
\label{subsec:lifecycle}

\begin{figure}[t]
\centering
\includegraphics[width=\columnwidth]{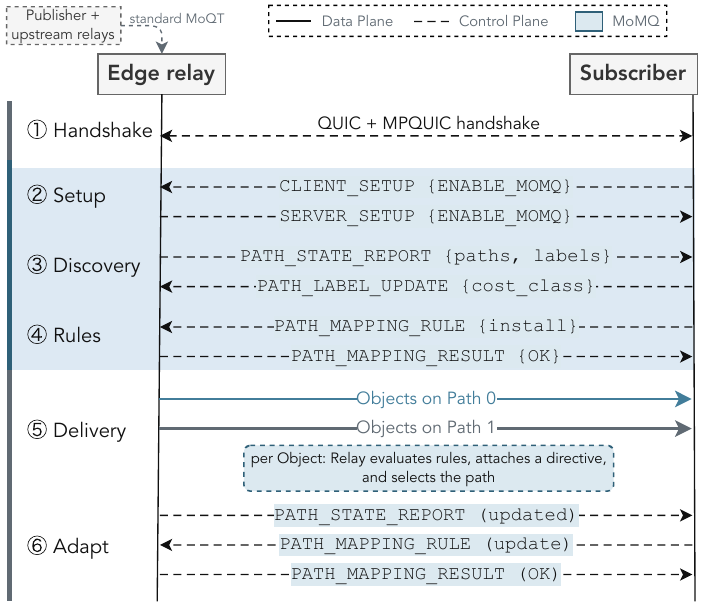}
\caption{A \momq session from handshake to steady state. The shaded exchanges are the only additions \momq makes, all control-plane, at setup and whenever paths or policy change. Delivery itself carries no \momq signaling, and upstream relays and the publisher see standard \moqt throughout.}
\label{fig:momq_lifecycle}
\end{figure}

\afblock{Protocol surface and lifecycle.}
\momq adds five wire-level elements to \moqt, one setup parameter and four control messages, and \cref{fig:momq_lifecycle} shows how a session uses them.
\Cref{app:format} gives the byte layout and code points of each element.
Both endpoints exchange \texttt{ENABLE\_MOMQ} in the \moqt setup, after which the relay reports its paths and labels in a \texttt{PATH\_STATE\_REPORT} and the subscriber annotates them through \texttt{PATH\_LABEL\_UPDATE}.
The subscriber then installs rules with \texttt{PATH\_MAPPING\_RULE}, and the relay accepts or rejects each one with \texttt{PATH\_MAPPING\_RESULT}.
Those four messages are the whole of the control plane.
The channel carries no compliance feedback, since \texttt{PATH\_MAPPING\_RESULT} acknowledges that a rule was installed rather than that it was followed.
A subscriber verifies that its preferences take effect by observing delivery.
Steady-state delivery then runs the scheduling pipeline on relay-local state alone, and no further \momq message is exchanged until a path changes.
When one does, a fresh \texttt{PATH\_STATE\_REPORT} lets the subscriber adjust rules or labels without interrupting delivery.

\momq state is session-scoped and isolated between subscribers.
A session in which either side omits \texttt{ENABLE\_MOMQ} carries no \momq messages and behaves as standard \moqt.
Note that we do not change anything else in \moqt.
Group ordering, Object status, caching, priority scheme and the stream-mapping modes all behave as before.
Subscriber and publisher priority continue to govern which Objects the relay forwards first at the \moqt layer, while a \momq \texttt{PRIORITY} directive acts one layer down and orders the streams already released to the transport.

\section{\momq in Practice}
\label{sec:casestudy}
\label{sec:prototype}

We implement \moqt and its \momq extension in Rust on top of TQUIC~\cite{tquic}, an open-source QUIC stack with \mpquic support.
The prototype exists to test whether rule-driven path selection produces the behavior the design predicts, not to serve production traffic.
Three binaries form the delivery chain, where the publisher encodes media into GOP-structured Objects, the relay forwards them and hosts the \momq rule path, and the subscriber reconstructs frames and records per-frame delivery timing.

The division between reused and added code follows the protocol layering.
The QUIC connection machinery, \mpquic path management, and per-path congestion control are used unchanged, with each path's controller assigned by static configuration.
In the \moqt layer, the Object encoding gains the metadata block and nothing else, so every base header field keeps its wire format.
\momq's setup parameter and four control messages live in a protocol module separate from the base \moqt message set, so the extension is additive in code as it is on the wire.
Changes to the transport itself are confined to its scheduler, which is extended to read a per-stream directive during path selection and to admit prioritized data into the slack of a multi-round transmission (\cref{subsec:svc_rules}).

Installed rules live in a per-session store that rule evaluation consults for every forwarded Object.
Installing a rule under an existing identifier replaces it. 
\momq and \mpquic are negotiated independently, so a subscriber can install rules while only one path exists and the policy takes effect without an additional round trip when a second path appears.
Metadata is attached as UTF-8 keys with byte-string values, so frame type, temporal layer, and dependency identifiers are present for each Object.
The relay reads these fields for matching and never parses the payload, which remains encrypted end to end.
The metadata overhead is minimal, roughly 40--80~bytes on Objects whose payloads span 5--230~KB.
Note that our prototype does not implement multi-subscriber rule isolation, relay-side authorization of rule installation, or the resource limits as we consider them out of scope.
\cref{sec:discussion} discusses how new rule sets could reach a relay.

\afblock{Crossing the layer boundary.}
\label{subsec:impl_directive}
When rule evaluation completes, the relay binds the merged directive to the QUIC stream that carries the Object, as implementation-local context that the scheduler can read.
Before attachment, a QUIC stream is an opaque byte channel that the scheduler treats like any other.
After attachment, the scheduler reads the directive from the stream without parsing any \moqt framing.
\moqt carries each Subgroup on its own QUIC stream, so a stream corresponds to a temporal layer.
Many Objects therefore share one stream, and the relay refreshes the stream's directive with each Object it forwards.
When the scheduler next selects a path, it reads the directive of every stream that is ready to send.
Each dimension acts as \cref{subsec:rules} defines it, so the priority sets the dequeue order under contention, the balancing constraint gates cross-path splitting, the preference biases the scheduler's path scoring, and the affinity resolves through the bounded history of recent assignments.
If no rule matches, the directive holds defaults, namely priority zero, \texttt{SINGLE\_PATH}, no preference, and no affinity (\cref{subsec:pipeline}).

Suppose an Object arrives carrying \texttt{frame\_type=IDR}, and an illustrative rule assigns matching Objects priority 100, \texttt{SINGLE\_PATH}, and a preference for \texttt{cost\_class=free}.
Rule evaluation produces exactly that directive, directive attachment binds it to the Object's stream, and path selection weighs it against two live paths, a 45~ms satellite path labeled \texttt{cost\_class=free} and a 15~ms metered path.
The scheduler dequeues this stream first, keeps the frame whole, and sends it on the free path even though the metered path is 30~ms faster (enforced by the cost-sensitive behavior rule in \cref{subsec:svc_rules}).

Information also crosses this boundary in the other direction, and again only as reads.
Rule-side state may consult transport state, as the interleave budget of \cref{subsec:svc_rules} consults a congestion-window estimate, but no rule can set a congestion window, retransmit a packet, or pin a path.
Labels are the only channel through which subscriber input could reach transport configuration, so the implementation bounds this channel as well.
Descriptive labels take effect only through rules that reference them, and the subscriber-over-relay label merge of \cref{subsec:metadata} applies to descriptive keys alone.
An implementation may treat selected keys as operational, for instance a key that selects a path's congestion controller, but such keys are accepted only from operator configuration against an allowlist.
The relay also rate-limits its own \texttt{PATH\_STATE\_REPORT} messages to one per smoothed RTT, since path state can change on millisecond timescales during a reconfiguration and per-fluctuation reports would be stale before the subscriber could act on them.

\subsection{Case Study: \momq over LEO-Terrestrial Networks}
\label{subsec:svc_rules}

We build a complete specialization, an SVC conferencing policy for the Starlink and WiFi path pair that \cref{subsec:eval_setup} describes.
Recall from \cref{subsec:pipeline} that a specialization has two halves, a label provider that annotates paths from local knowledge and rules that reference the annotation.
For this pairing, the local knowledge is the satellite path's reconfiguration schedule, which the label provider tracks, and each path's cost class, which static labels record.
The same derivation applies to a cellular pairing, where the literature already provides congestion control tuned to wireless access~\cite{zhuge} and path preferences for metered cellular links~\cite{mp-dash}.

\Cref{tab:svc_rule_set} shows the rules, \rone through \rfour, along with the motivating findings F1 through F5 from our experiments on testbed nodes equipped with a Starlink and a terrestrial connection.
\label{subsec:reconf_char}
\Cref{fig:reconf_analysis} shows how the single-path failures of \cref{subsec:four_problems} appear on the satellite path.
In \cref{fig:reconf_analysis}a, nearly every frame delayed past 400~ms sits inside a reconfiguration window (F2).
\Cref{fig:reconf_analysis}b shows the convoy behind each \iframe transmission.
The \iframe carries the highest delayed-frame ratio at 14.8\%, and the following \pframes decay from 7.2\% toward 1\%.
Frames before it stay near 0.9\%, so the blocking runs forward from the \iframe and never behind it (F1).
The disruption itself is schedulable, because the 15-second reconfiguration cycle of \cref{subsec:four_problems} runs on a constellation-wide wall clock, at the 12th, 27th, 42nd, and 57th second of each minute~\cite{tanveer-constellations, mohan-starlink, dissectstarlink2026sigcomm}.
Our measurements confirm this schedule from a separate vantage point, with every observed interval within $\pm$50~ms of its predicted time, so the timing is predictable enough to schedule against (F3).
The events are also brief enough that a short avoidance window suffices, with a median duration of 58~ms and 88\% completing within 100~ms (F4, with the full distribution in \cref{app:reconf_duration}).

\begin{figure}[t]
\centering
\includegraphics[width=\columnwidth]{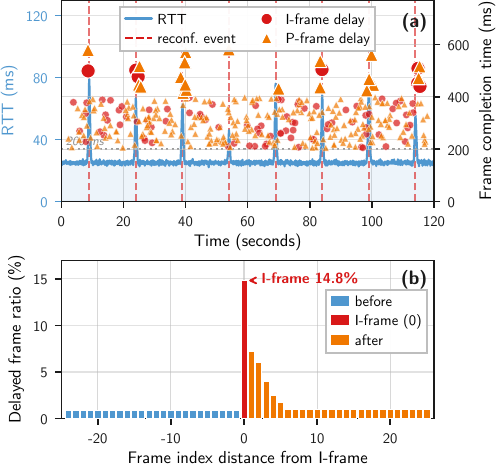}
\caption{Delayed frames on Starlink. (a)~Frame completion times beyond 200~ms over the sender-perceived RTT trace, with reconfiguration events marked and delays beyond 400~ms drawn as larger markers. (b)~Delayed-frame ratio by frame index distance from the \iframe across 100~GOPs.}
\label{fig:reconf_analysis}
\end{figure}

\afblock{\rone Budget-based \pframe interleaving} addresses the convoy effect of \cref{subsec:four_problems} (F1).
During an \iframe transmission spanning $k$ RTT cycles there is slack capacity, namely the difference between what the congestion window carries across those cycles and the \iframe size, given by
\begin{equation}
B_{\text{budget}} = \left\lceil \frac{S_{I}}{\mathit{cwnd}_{\text{cons}}} \right\rceil \cdot \mathit{cwnd}_{\text{cons}} - S_{I}
\label{eq:interleave_budget}
\end{equation}
where $k = \lceil S_{I} / \mathit{cwnd}_{\text{cons}} \rceil$.
This slack carries \pframe packets without adding a transmission round, since the \iframe needs $k$ rounds whether or not the unused capacity is filled.
\Cref{subsec:eval_mechanism} confirms that \iframe completion is unchanged when interleaving is active.
We derive $\mathit{cwnd}_{\text{cons}} = \widehat{BW}_{m} \times bRTT_{l}$ from LeoCC's conservative Kalman-filtered bandwidth and minimum RTT estimates~\cite{leocc}, since an overestimate would admit \pframe packets that compete with the \iframe instead of filling its slack.
For a typical case with $S_{I} = 230$~KB and $\mathit{cwnd}_{\text{cons}} = 80$~KB, this yields $k = 3$ rounds and $B_{\text{budget}} = 10$~KB, roughly one \pframe per \iframe transmission.
The action of \rone is a plain priority grant, because interleaving is not an action type.
A relay-side budget tracker admits a prioritized \pframe into the current congestion-window round only while $B_{\text{budget}}$ has room, and a zero budget admits nothing, so the mechanism cannot make things worse.

\afblock{\rtwo Reconfiguration-aware scheduling} addresses the reconfiguration collision of \cref{subsec:four_problems}, which dominates the tail (F2).
Dropping the affected frame is no escape, because a lost \iframe invalidates its entire GOP, so delivery must be ensured rather than abandoned (F5).
Rather than building a prediction model, we exploit the deterministic schedule directly.
The label provider of this specialization is a reconfiguration marker.
During negotiation the subscriber labels the satellite path, and from then on the marker injects \texttt{leo\_state=reconf} into the match context during a danger zone of $\pm$100~ms around each scheduled reconfiguration time, on its own clock and with no per-event signaling.
While the annotation is present, \iframes prefer the backup path, and \pframes continue on the satellite path, where a disrupted frame costs one retransmission delay rather than a stalled GOP.
The window is conservative against the measured start jitter and the typical event duration.
Because four danger zones per minute occupy 800~ms in total, or 1.3\% of time, the additional backup-path usage is small.

\afblock{\rthree Dependency-aware co-location} addresses the dependency violation of \cref{subsec:four_problems}.
The mechanism routes each frame to the same path as the frame it depends on, using a bounded history of recent Object-to-path assignments.
No dependency crosses a GOP boundary, so a history of one GOP's worth of entries covers every reference the rule can encounter.
If the referenced Object is absent from the history, or its recorded path has become unavailable, the frame falls through to default selection, so a stale history degrades placement rather than delivery.

\afblock{\rfour Cost-sensitive path preference} addresses the backup overuse of \cref{subsec:four_problems}.
Unlike the first three mechanisms, it encodes a cost constraint rather than a measured timing failure.
The mechanism labels each path with a \texttt{cost\_class} attribute and defaults to the cost-free path unless a higher-priority rule overrides it.
The latency-oriented schedulers of \cref{subsec:mpquic_fails} treat the metered path as general-purpose overflow.
Under this rule set it carries only what a higher-priority rule sends there, chiefly \iframes inside danger zones.

Two of the rule set's match keys are not publisher metadata.
\texttt{pending\_iframe} and \texttt{leo\_state} are the relay annotations that the match model of \cref{subsec:rules} admits, injected by the budget tracker while an \iframe is in flight and by the reconfiguration marker during danger zones.
The relay remains application-agnostic, because the annotations describe conditions the relay observes or computes locally, and the application's own keys stay opaque.

\begin{table}[t]
\centering
\caption{The four \momq rules for SVC conferencing, in descending priority order. Each rule closes one delivery failure of \cref{subsec:four_problems}.
}
\label{tab:svc_rule_set}
\vspace*{-1em}
\small
\setlength{\tabcolsep}{3pt}
\begin{tabularx}{\columnwidth}{@{}c >{\hsize=1.12\hsize}L >{\hsize=1.28\hsize}L >{\hsize=0.60\hsize}L@{}}
\toprule
\textbf{\#} & \textbf{Match} & \textbf{Action} & \textbf{Closes} \\
\midrule
\rone & \texttt{frame\_type=P} and \texttt{pending\_iframe=}\allowbreak\texttt{true} & PRIORITY (highest) & Convoy effect \\
\rtwo & \texttt{frame\_type=IDR} and \texttt{leo\_state=}\allowbreak\texttt{reconf} & PATH\_PREFERENCE \texttt{cost\_class=}\allowbreak\texttt{metered} & Reconf.\ collision \\
\rthree & \texttt{depends\_on} EXISTS & PATH\_AFFINITY & Dependency violation \\
\rfour & catch-all & PATH\_PREFERENCE \texttt{cost\_class=}\allowbreak\texttt{free} & Backup overuse \\
\bottomrule
\end{tabularx}
\end{table}

Rule composition follows the per-dimension merge of \cref{subsec:rules}.
An enhancement-layer \pframe arriving during a danger zone while an \iframe is in flight matches \rone, \rthree, and \rfour, whose priority, affinity, and preference merge into its directive.
The priority decides when the frame is dequeued and the affinity decides which path it takes, at once rather than in sequence.
An \iframe in the same danger zone exercises the preference merge instead, where the higher rule priority of \rtwo overrides the catch-all preference of \rfour.
Every dimension of the outcome is resolved by exactly one rule, and only the scheduler's advisory override remains discretionary.
The policy uses three of the four action types directly, and the fourth, \texttt{BALANCING}, acts through its conservative \texttt{SINGLE\_PATH} default.

\section{Evaluation}
\label{sec:evaluation}

\subsection{Experimental Setup}
\label{subsec:eval_setup}

Our testbed consists of a Starlink terminal alongside a university campus WiFi connection in Europe, connected to a server in a commercial datacenter in Germany, roughly 700~km away.
The Starlink terminal provides LEO satellite connectivity with throughput between 50 and 200~Mbps and RTT between 25 and 80~ms depending on satellite position and reconfiguration state.
The WiFi connection is the secondary path and emulates a metered cellular backup link, so traffic on it carries a per-byte cost.
We vary this topology at both ends (\cref{subsec:eval_diversity}).
A second relay runs in a commercial datacenter in Finland, and a second subscriber site in western Canada holds its own satellite and terrestrial path pair.

We transmit a 100-second 1080p video encoded with SVC~\cite{svc-overview} using 1-second GOPs at 50~fps.
The encoding uses the three-temporal-layer structure described in \cref{subsec:bg_svc}, with each GOP anchored by a single \iframe, and \cref{app:encoding} tabulates the per-frame-type sizes at this operating point.
The choice of 1-second GOPs reflects common conferencing practice.
Shorter GOPs recover faster from loss but raise bitrate through more frequent \iframes, while longer GOPs are more efficient but amplify the impact of each \iframe loss.

We compare seven configurations over these two paths.
Two single-path baselines carry \moqt over regular QUIC on one link alone.
Four transport-only \mpquic configurations run the state-of-the-art popular schedulers, namely Round-Robin, MinRTT, BLEST, and Redundant, which together span the byte-level design space from cyclic distribution and lowest-RTT preference to blocking avoidance and full duplication.
\momq installs the four rules of \cref{subsec:svc_rules}, so the only difference between \momq and the transport-only configurations is that Object metadata reaches the path decision.

We test four congestion control algorithms, namely Cubic~\cite{cubic}, BBRv3~\cite{bbrv3}, Copa~\cite{copa}, and LeoCC~\cite{leocc}, the last only on Starlink since it is designed for LEO links.
In multipath configurations each path runs its own controller, LeoCC on Starlink and Cubic on WiFi (\cref{sec:prototype}).
Because the relay terminates the upstream \moqt session rather than tunneling it, these controllers run in series with the publisher's own control loop rather than nested inside it, so the interaction that degrades tunneled multipath transports does not arise~\cite{pieska-nested}.
The single-path baselines use the best controller for each link, Copa on WiFi and LeoCC on Starlink, which \cref{app:cca_baselines} establishes by comparing all four controllers per link.
All measurements record per-packet RTT timestamps at millisecond precision.

\emph{Frame completion time} (FCT), the duration from first packet transmission to last packet acknowledgment of a frame, captures delivery latency at the transport level.
\emph{Minimum buffer length}, the smallest playback buffer ensuring stutter-free playback, captures the quality of experience at the receiver.
\emph{Backup-path usage}, the fraction of traffic carried by the metered path, captures the cost of a configuration.
Each configuration is measured over 10~runs of the same source video, yielding 100~buffer datapoints and $\approx$5\,000 frames per configuration.
In addition to means, we also report percentiles since a small number of extreme events, such as a single \iframe and reconfiguration collision, can dominate user experience while leaving the mean nearly unchanged.
Each topology still measures one subscriber site with one satellite and terrestrial path pair, so the absolute numbers are specific to that vantage point.
\Cref{subsec:eval_diversity} compares across the relay and subscriber variations to establish which of the conclusions are generalizable.

\subsection{Performance Results}
\label{subsec:eval_results}

\afblock{Transport-level performance.}
\label{subsec:eval_transport}
\label{subsec:eval_headline}
\Cref{tab:tail_comparison} reports frame completion time, playback buffer, and backup-path usage for all seven configurations through the Germany relay.
\Cref{fig:mpquic_comparison}a plots the buffer distributions and \cref{fig:mpquic_comparison}b the FCT percentiles.
Even the stronger of the two single-path baselines, Copa over WiFi, misses the 150~ms interactive target~\cite{itu-g114} by more than a factor of two on both tail FCT and playback buffer.
LeoCC on Starlink trails further, because every reconfiguration lands on its only path and stretches its worst tails past half a second.

Transport-only multipath does not close this gap.
In \cref{fig:mpquic_comparison}a, every scheduler's buffer latency sits above the 150~ms target, and none improves meaningfully on single-path WiFi.
MinRTT routes $\approx$90\% of traffic to the lower-RTT WiFi path, and when a 230~KB \iframe fills the 82~KB WiFi congestion window, the overflow packets spill onto Starlink at 45~ms RTT.
The receiver then waits for these slower packets before the \iframe can be decoded, and the resulting cross-path head-of-line blocking leaves MinRTT behind single-path WiFi on every metric.
Round-Robin distributes each \iframe's packets cyclically, so about half arrive via WiFi at 15~ms while the rest arrive via Starlink at 45~ms, and the systematic 30~ms per-round reordering penalty accumulates across three or more rounds.
It is the worst configuration overall.
BLEST avoids Starlink whenever it estimates blocking would occur, which makes it in practice a smarter single-path WiFi scheduler, and its small edge comes at $\approx$92\% usage of the metered path.
Redundant duplicates every packet on both paths and matches single-path WiFi at twice the bandwidth cost.
The residual tails of the four schedulers share one cause, an \iframe in flight when the satellite path reconfigures (\cref{subsec:four_problems}).

In \cref{fig:mpquic_comparison}b, every baseline crosses the 150~ms target between its median and its P99.
\momq reaches a P99.9 of 114.1~ms, a 70.3\% reduction from the best transport-only scheduler (BLEST) and 71.5\% from the best single-path baseline (Copa over WiFi), and its playback buffer undercuts the best baseline by 61.5\%.
It is the only configuration that meets the interactive target on both tail latency and buffer, and it does so with 11.3\% of traffic on the metered backup against 90--100\% for the latency-oriented transport schedulers.
Its median FCT sits a few milliseconds above BLEST's, the cost of keeping bulk traffic on the flat-rate satellite path rather than the faster metered one.
\ul{All seven configurations hold the same two paths and the same per-path controllers, so the 70.3\% tail reduction comes from \momq mapping Objects to paths by their metadata.}

\begin{figure}[t]
    \centering
    \includegraphics[width=\columnwidth]{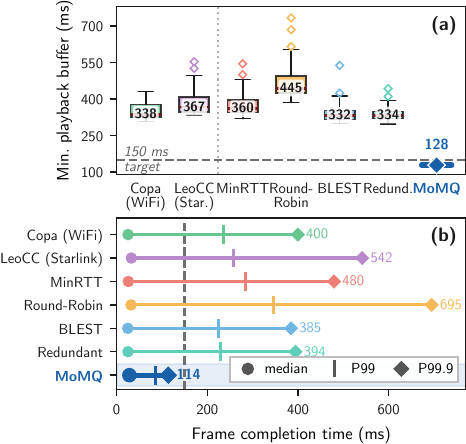}
    \caption{\momq against single-path and transport-only multipath baselines. (a)~Minimum playback buffer with the 150~ms interactive target marked. (b)~Frame completion time for all configurations as median, P99, and P99.9.}
    \label{fig:mpquic_comparison}
    \end{figure}

\afblock{Contribution of scheduling rules.}
\label{subsec:eval_rules}
\label{subsec:eval_mechanism}
\label{subsec:eval_ablation}
Activating the four rules of \cref{subsec:svc_rules} incrementally on the live testbed lowers the median buffer from $\approx$248~ms under the cost-sensitive default (\rfour) alone to $\approx$128~ms with the full set, while backup-path usage grows only from 0.3\% to the 11.3\% reported above.
The activation steps add interleaving (\rone), reconfiguration avoidance (\rtwo), and dependency co-location (\rthree) in turn.
\Cref{app:ablation_live} tabulates each activation step, and a controlled simulation confirms that the contributions are statistically robust and near-additive, with all pairwise comparisons significant at $p < 0.001$ (\cref{app:ablation_sim}).

\begin{table}[tb]
    \centering
    \caption{Frame completion time, playback buffer, and backup-path usage for the seven configurations through the Germany relay.}
    \label{tab:tail_comparison}
    \small
    \begin{tabularx}{\columnwidth}{@{}L c c c c c@{}}
    \toprule
    \textbf{Configuration} & \textbf{Med.} & \textbf{P99} & \textbf{P99.9} & \textbf{Buf.} & \textbf{Backup} \\
     & \textbf{(ms)} & \textbf{(ms)} & \textbf{(ms)} & \textbf{(ms)} & \textbf{use} \\
    \midrule
    Copa / WiFi & 25.2 & 235.0 & 399.8 & 338.5 & 100\% \\
    LeoCC / Starlink & 32.6 & 258.4 & 541.7 & 367.6 & 0\% \\
    MPQUIC MinRTT & 25.9 & 284.6 & 479.5 & 360.2 & 89.7\% \\
    MPQUIC Round-Robin & 31.8 & 345.2 & 694.9 & 445.3 & 47.6\% \\
    MPQUIC BLEST & 24.8 & 225.1 & 384.7 & 332.4 & 91.8\% \\
    MPQUIC Redundant & 25.1 & 229.7 & 394.5 & 334.2 & 100\% \\
    \rowcolor{gray!15}
    \multicolumn{1}{@{}>{\columncolor{gray!15}[0pt][\tabcolsep]}l}{\textbf{\momq}} & \textbf{28.2} & \textbf{85.6} & \textbf{114.1} & \textbf{127.9} & \multicolumn{1}{>{\columncolor{gray!15}[\tabcolsep][0pt]}c@{}}{\textbf{11.3\%}} \\
    \bottomrule
    \end{tabularx}
    \end{table}

To locate each rule's effect, \cref{fig:fct_cdf} decomposes per-frame completion time by frame type across an extended measurement campaign.
In \cref{fig:fct_cdf}a, the three configurations that enable WiFi offloading show a distinct step near 8--14~ms, corresponding to \pframes routed via WiFi with a one-way delay of about 8~ms.
The cost-sensitive default, which confines all traffic to Starlink, produces no such step and begins near 25~ms.
The inset of \cref{fig:fct_cdf}a magnifies the tail above P97, where the four configurations split into two clusters at P99.9.
The full system and the reconfiguration-avoidance configuration form a tight cluster near 110~ms, while cost-sensitive and interleaving remain above 217~ms, so reconfiguration avoidance alone removes roughly 100~ms of tail latency.
Without it, an \iframe and reconfiguration collision sets the P99.9 of the whole distribution.
\Cref{fig:fct_cdf}b isolates \iframe FCT.
Cost-sensitive and interleaving have nearly identical \iframe distributions, which confirms that \pframe interleaving does not affect \iframe scheduling.
Interleaved \pframes travel in the slack capacity of the \iframe's congestion-window rounds without altering the \iframe's own path assignment.
Separation begins with reconfiguration avoidance, whose rerouting during danger zones sends 6.7\% of \iframes over WiFi and produces the low-FCT step in the 16--50~ms range.
This compresses \iframe P99 from 286~ms to 119~ms, since a rerouted \iframe avoids both the outage and the congestion-window collapse that follows it.
Full \momq trims a further 5~ms through dependency co-location.

\Cref{fig:fct_cdf}c shows that \pframe tail compression drives the buffer improvements.
Under cost-sensitive scheduling, \pframes have a P99/P50 ratio of $3.5\times$, driven by head-of-line blocking while multi-round \iframes are transmitted.
Interleaving cuts \pframe P99 from 122~ms to 54~ms.
Reconfiguration avoidance compresses it further to 43~ms by eliminating the cascading effect of collisions on subsequent \pframes.
The last activation step, dependency co-location, brings the ratio to $1.2\times$, so almost all \pframes complete in a single congestion-window round without queueing behind larger frames.
\ul{Interleaving cuts P-frame P99 by 56\% and reconfiguration avoidance cuts I-frame P99 by 58\%, each on a frame type the other leaves unchanged, so both rules are needed to meet the interactive target.}

\afblock{Path diversity and relay location.}
\label{subsec:eval_diversity}
\label{subsec:eval_location}
The evaluation so far uses the Germany relay.
We vary the topology in two directions, increasing the relay distance and moving the subscriber to a site whose two paths converge, and find that distance leaves the gains intact while path diversity is indispensable.
Both variations move exactly one endpoint of the multipath segment where \momq operates (\cref{subsec:architecture}), while the publisher and its single-path hop toward the relay stay fixed, so no new path elsewhere in the delivery chain can influence the comparison.

We first move the relay to Finland, roughly 1\,800~km from the subscriber and $2.5\times$ the original distance, and verify with traceroute beforehand that the WiFi and Starlink paths do not share routers.
The two paths remain distinct through separate transit networks all the way to the relay, with one-way delays of approximately 30~ms via WiFi and 50~ms via Starlink.
We repeat the comparison against the two single-path baselines and BLEST, the strongest transport-only scheduler in \cref{tab:tail_comparison}.
\momq reduces P99.9 by 52--70\%, needs less than half the playback buffer of any baseline, and again leaves the metered path nearly idle at 9.3\% (\cref{tab:finland_comparison}).
\momq finishes nearly all \iframes below 150~ms, while all three baselines extend to 400--550~ms (\cref{app:finland_cdf} shows the full distributions).
The \pframe median of \momq again sits above BLEST, since bulk traffic stays on the flat-rate satellite path, whose one-way delay exceeds that of the metered path by 20~ms.
The rule contributions reproduce as well.
Interleaving (\rone) cuts the buffer from $\approx$284~ms under the cost-sensitive default to $\approx$235~ms, reconfiguration avoidance (\rtwo) brings it to $\approx$175~ms, and the full rule set reaches $\approx$164~ms.
The policy carries over unmodified, since its rules reference metadata fields and path labels rather than absolute latency thresholds, and the configuration ranking is unchanged.

\begin{figure}[!tb]
    \centering
    \includegraphics[width=\columnwidth]{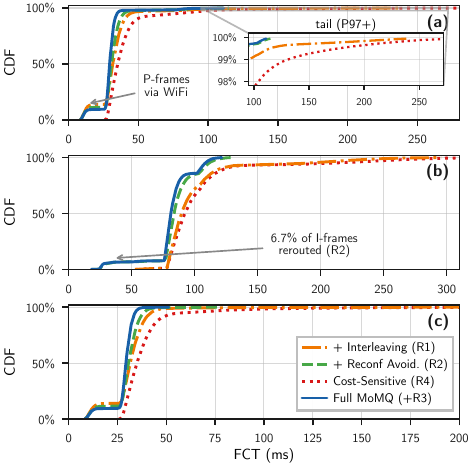}
    \caption{Frame completion time under incremental rule activation (250~runs and $\approx$750\,000~frames per configuration). (a)~All frames (P97 magnified). (b)~\iframe FCT. (c)~\pframe FCT.}
    \label{fig:fct_cdf}
    \end{figure}

\begin{table}[tb]
\centering
\caption{Baseline comparison with the relay in Finland, at $2.5\times$ the original relay distance. \pframe and \iframe medians are in ms.}
\label{tab:finland_comparison}
\small
\begin{tabularx}{\columnwidth}{@{}L c c c c c@{}}
\toprule
\textbf{Configuration} & \textbf{PF} & \textbf{IF} & \textbf{P99.9} & \textbf{Buf.} & \textbf{Backup} \\
 & \textbf{med.} & \textbf{med.} & \textbf{(ms)} & \textbf{(ms)} & \textbf{use} \\
\midrule
Copa / WiFi & 34.3 & 142.3 & 364.5 & 396.6 & 100.0\% \\
LeoCC / Starlink & 53.5 & 186.6 & 450.2 & 440.1 & 0.0\% \\
MPQUIC BLEST & 33.7 & 132.9 & 275.3 & 371.5 & 84.3\% \\
\rowcolor{gray!15}
\multicolumn{1}{@{}>{\columncolor{gray!15}[0pt][\tabcolsep]}l}{\textbf{\momq}} & \textbf{45.1} & \textbf{95.9} & \textbf{132.4} & \textbf{164.3} & \multicolumn{1}{>{\columncolor{gray!15}[\tabcolsep][0pt]}c@{}}{\textbf{9.3\%}} \\
\bottomrule
\end{tabularx}
\end{table}

We then move the subscriber to western Canada, roughly 6\,000~km from the relay, with the subscriber again connected via both Starlink and a terrestrial path.
The relay stays in Germany, since relocating it near the subscriber would replace the upstream delivery path rather than stretch the subscriber-to-relay segment whose path diversity is under test.
The advantage nearly disappears.
\momq improves on BLEST by only 5.0\% in median FCT and 6.6\% in median buffer, far below the roughly 70\% reductions with the Germany relay.

\Cref{tab:canada_traceroute} in \cref{app:traceroute} aligns traces hop by hop and confirms that the two paths converge onto a shared backbone at Chicago, so the subscriber sees only one path to the relay.
From there both paths follow the same backbone sequence through Newark, London, and Hamburg, and a repeated MPLS label on those hops supports a shared core segment.
The traces rejoin inside the destination datacenter before the target host.
Once traffic enters this shared long-haul segment, both paths experience nearly the same queueing and propagation conditions, so multipath scheduling has little leverage.
Reconfiguration avoidance (\rtwo) contributes little because the shared core tail masks reconfiguration events, and \pframe interleaving (\rone) gains nothing when both paths deliver frames at comparable latencies.
\momq's rules assume path diversity that does not exist on this topology.

\Cref{fig:location_comparison} tracks the median playback buffer of \momq and BLEST across the three topologies, where \momq stays near the interactive target through Germany and Finland and then converges with BLEST at Canada.
Moving the relay from Germany to Finland raises \pframe medians by 55--94\%, because added propagation delay has a larger relative effect on small, fast frames.
\iframe medians, tails, and buffers rise far less, and \momq stays ahead on every metric in both locations.
\Cref{fig:location_cdf} overlays the full FCT distributions for Finland and Canada.
The Finland curves cross, where BLEST holds a slightly lower median and \momq a much shorter tail, whereas the Canada curves track each other through the body of the distribution once the paths converge.
\ul{The gains survive a $2.5\times$ increase in relay distance without rule changes and fall to 5--7\% once the paths share a backbone, so path diversity, not relay proximity, is the deployment precondition.}

\begin{figure}[t]
\centering
\begin{subfigure}{0.49\columnwidth}
\centering
\includegraphics[width=\linewidth]{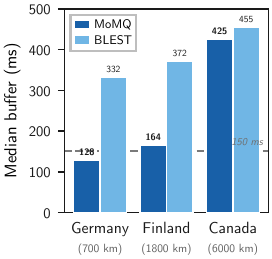}
\caption{Median playback buffer}
\label{fig:location_comparison}
\end{subfigure}
\hfill
\begin{subfigure}{0.49\columnwidth}
\centering
\includegraphics[width=\linewidth]{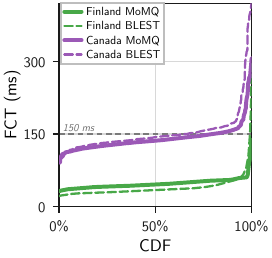}
\caption{Frame completion time}
\label{fig:location_cdf}
\end{subfigure}
\caption{\momq against BLEST across relay and subscriber distance.
(a)~Median playback buffer at each location, with the 150~ms interactive target marked.
(b)~All-frame FCT distributions for the Finland and Canada topologies, where color marks location and line style marks configuration.}
\label{fig:location_topologies}
\end{figure}

\section{Discussion and Conclusion}
\label{sec:discussion}

Multipath transport gives endpoints a choice of paths, and \moqt gives relays named Objects with metadata.
We presented \momq, which connects the two by letting an application install declarative object-to-path mapping rules at the relay.
The rules match metadata the relay never interprets and express delivery preferences against abstract path labels, so the relay acts on media semantics while containing no media logic.
In a case study on SVC conferencing over Starlink and WiFi (\cref{sec:casestudy}), four rules built from two match operators reduced P99.9 frame completion time from $\approx$385~ms under the best transport-only scheduler to $\approx$114~ms (\cref{subsec:eval_headline}).
No other configuration met the 150~ms interactive budget.
The same rules cut metered-path usage from $\approx$92\% to $\approx$11\%, and they retained their advantage without retuning at $2.5\times$ the relay distance.
The measured characterization of Starlink's reconfiguration timing and duration (\cref{subsec:reconf_char}) also provides a reference for protocol work over LEO links.

The results also mark the design's boundaries, and each suggests an extension.
(i)~The gains require genuinely disjoint paths, since the advantage fell from $\approx$70\% to $\approx$5\% when the Canada subscriber's paths converged onto a shared backbone.
A subscriber cannot see such convergence from local interface state, so path-disjointness detection would let a relay decline rules that cannot improve delivery.
(ii)~\momq schedules only as well as the installed rules, which the relay validates syntactically rather than semantically.
Nothing in the design, however, fixes where a rule set originates.
A CDN operator could curate rule sets with its customers, a session could instantiate a rule set shipped as a transport add-on, or automated generation from application-declared objectives could remove the authoring burden.
(iii)~Relays are assumed cooperative.
Advisory semantics accommodate benign deviation, where a relay overrides a rule under resource pressure, but the protocol neither detects nor prevents malicious deviation.
The metadata and rules a relay receives also reveal the application's encoding structure and the subscriber's cost preferences.
The evaluation is similarly bounded to one constellation, one SVC operating point, and a single subscriber per relay.
Different codecs, constellations, and production loads would shift the value of individual rules, though not the interface that expresses them.
(iv)~\momq schedules only the subscriber's last hop (\cref{sec:design}), and it assumes both paths meet at a single relay.
A mobile publisher, such as a news broadcast van uplinking over cellular and satellite, holds the same path diversity toward its first relay, so the rule interface could extend upstream to cover the delivery chain end to end.
A subscriber whose access networks reach different relays would instead need rules that span relays, where path selection becomes relay selection.
(v)~\momq steers Objects the encoder has already produced and leaves the encoding unchanged.
The path state a subscriber receives through \texttt{PATH\_STATE\_REPORT} messages could equally drive Track selection, so one control loop would carry both delivery steering and bitrate adaptation.

The broadest extension is beyond video.
The rule derivation of the case study (\cref{subsec:svc_rules}) applies wherever an application has heterogeneous data importance, inter-object dependencies, and cost-asymmetric paths.
A cloud gaming service can mark frames rendered in response to player input, a conference bridge the current speaker's Objects, and a live broadcast the featured camera's feed.
Each policy amounts to a label vocabulary and a few rules over the same five wire elements, and in each case the application supplies the domain knowledge while the relay learns none of it.
This design opens the question of how many applications' delivery policies become this small once the network offers a place to express them.

\bibliographystyle{ACM-Reference-Format}
\bibliography{references}


\begin{thebibliography}{65}


\ifx \showCODEN    \undefined \def \showCODEN     #1{\unskip}     \fi
\ifx \showISBNx    \undefined \def \showISBNx     #1{\unskip}     \fi
\ifx \showISBNxiii \undefined \def \showISBNxiii  #1{\unskip}     \fi
\ifx \showISSN     \undefined \def \showISSN      #1{\unskip}     \fi
\ifx \showLCCN     \undefined \def \showLCCN      #1{\unskip}     \fi
\ifx \shownote     \undefined \def \shownote      #1{#1}          \fi
\ifx \showarticletitle \undefined \def \showarticletitle #1{#1}   \fi
\ifx \showURL      \undefined \def \showURL       {\relax}        \fi
\providecommand\bibfield[2]{#2}
\providecommand\bibinfo[2]{#2}
\providecommand\natexlab[1]{#1}
\providecommand\showeprint[2][]{arXiv:#2}

\bibitem[Allamsetty(2024)]%
        {jitsi-svc}
\bibfield{author}{\bibinfo{person}{Jaya Allamsetty}.} \bibinfo{year}{2024}\natexlab{}.
\newblock \bibinfo{title}{{AV1} and more\ldots{} how does {Jitsi Meet} pick video codecs?}
\newblock \bibinfo{howpublished}{Jitsi Blog, \url{https://jitsi.org/blog/av1-and-more-how-does-jitsi-meet-pick-video-codecs/}}.
\newblock
\newblock
\shownote{Accessed August 2026}.


\bibitem[Arun and Balakrishnan(2018)]%
        {copa}
\bibfield{author}{\bibinfo{person}{Venkat Arun} {and} \bibinfo{person}{Hari Balakrishnan}.} \bibinfo{year}{2018}\natexlab{}.
\newblock \showarticletitle{Copa: Practical Delay-Based Congestion Control for the Internet}. In \bibinfo{booktitle}{\emph{15th USENIX Symposium on Networked Systems Design and Implementation (NSDI '18)}}. \bibinfo{publisher}{USENIX Association}, \bibinfo{address}{Renton, WA}, \bibinfo{pages}{329--342}.
\newblock


\bibitem[Aschenbrenner et~al\mbox{.}(2021)]%
        {mptcpScan-ifip}
\bibfield{author}{\bibinfo{person}{Florian Aschenbrenner}, \bibinfo{person}{Tanya Shreedhar}, \bibinfo{person}{Oliver Gasser}, \bibinfo{person}{Nitinder Mohan}, {and} \bibinfo{person}{J{\"o}rg Ott}.} \bibinfo{year}{2021}\natexlab{}.
\newblock \showarticletitle{From Single Lane to Highways: Analyzing the Adoption of Multipath TCP in the Internet}. In \bibinfo{booktitle}{\emph{2021 IFIP Networking Conference (IFIP Networking)}}. \bibinfo{pages}{1--9}.
\newblock
\href{https://doi.org/10.23919/IFIPNetworking52078.2021.9472785}{doi:\nolinkurl{10.23919/IFIPNetworking52078.2021.9472785}}


\bibitem[Bentaleb et~al\mbox{.}(2019)]%
        {bentaleb-abr}
\bibfield{author}{\bibinfo{person}{Abdelhak Bentaleb}, \bibinfo{person}{Bayan Taani}, \bibinfo{person}{Ali~C. Begen}, \bibinfo{person}{Christian Timmerer}, {and} \bibinfo{person}{Roger Zimmermann}.} \bibinfo{year}{2019}\natexlab{}.
\newblock \showarticletitle{A Survey on Bitrate Adaptation Schemes for Streaming Media Over {HTTP}}.
\newblock \bibinfo{journal}{\emph{IEEE Communications Surveys \& Tutorials}} \bibinfo{volume}{21}, \bibinfo{number}{1} (\bibinfo{year}{2019}), \bibinfo{pages}{562--585}.
\newblock
\href{https://doi.org/10.1109/COMST.2018.2862938}{doi:\nolinkurl{10.1109/COMST.2018.2862938}}


\bibitem[Blake et~al\mbox{.}(1998)]%
        {blake-diffserv}
\bibfield{author}{\bibinfo{person}{Steven Blake}, \bibinfo{person}{David Black}, \bibinfo{person}{Mark Carlson}, \bibinfo{person}{Elwyn Davies}, \bibinfo{person}{Zheng Wang}, {and} \bibinfo{person}{Walter Weiss}.} \bibinfo{year}{1998}\natexlab{}.
\newblock \bibinfo{booktitle}{\emph{An Architecture for Differentiated Services}}.
\newblock \bibinfo{type}{RFC} 2475. \bibinfo{institution}{Internet Engineering Task Force}.
\newblock
\href{https://doi.org/10.17487/RFC2475}{doi:\nolinkurl{10.17487/RFC2475}}


\bibitem[Bonaventure et~al\mbox{.}(2017)]%
        {rfc8041}
\bibfield{author}{\bibinfo{person}{Olivier Bonaventure}, \bibinfo{person}{Christoph Paasch}, {and} \bibinfo{person}{Gregory Detal}.} \bibinfo{year}{2017}\natexlab{}.
\newblock \bibinfo{booktitle}{\emph{Use Cases and Operational Experience with Multipath {TCP}}}.
\newblock \bibinfo{type}{RFC} 8041. \bibinfo{institution}{Internet Engineering Task Force}.
\newblock
\href{https://doi.org/10.17487/RFC8041}{doi:\nolinkurl{10.17487/RFC8041}}


\bibitem[Bosshart et~al\mbox{.}(2014)]%
        {p4}
\bibfield{author}{\bibinfo{person}{Pat Bosshart}, \bibinfo{person}{Dan Daly}, \bibinfo{person}{Glen Gibb}, \bibinfo{person}{Martin Izzard}, \bibinfo{person}{Nick McKeown}, \bibinfo{person}{Jennifer Rexford}, \bibinfo{person}{Cole Schlesinger}, \bibinfo{person}{Dan Talayco}, \bibinfo{person}{Amin Vahdat}, \bibinfo{person}{George Varghese}, {and} \bibinfo{person}{David Walker}.} \bibinfo{year}{2014}\natexlab{}.
\newblock \showarticletitle{{P4}: Programming Protocol-Independent Packet Processors}.
\newblock \bibinfo{journal}{\emph{ACM SIGCOMM Computer Communication Review}} \bibinfo{volume}{44}, \bibinfo{number}{3} (\bibinfo{year}{2014}), \bibinfo{pages}{87--95}.
\newblock
\href{https://doi.org/10.1145/2656877.2656890}{doi:\nolinkurl{10.1145/2656877.2656890}}


\bibitem[Cardwell et~al\mbox{.}(2023)]%
        {bbrv3}
\bibfield{author}{\bibinfo{person}{Neal Cardwell}, \bibinfo{person}{Yuchung Cheng}, \bibinfo{person}{Kevin Yang}, \bibinfo{person}{David Morley}, \bibinfo{person}{Soheil~Hassas Yeganeh}, \bibinfo{person}{Priyaranjan Jha}, \bibinfo{person}{Yousuk Seung}, \bibinfo{person}{Van Jacobson}, \bibinfo{person}{Ian Swett}, \bibinfo{person}{Bin Wu}, {and} \bibinfo{person}{Victor Vasiliev}.} \bibinfo{year}{2023}\natexlab{}.
\newblock \bibinfo{title}{{BBRv3}: Algorithm Bug Fixes and Public Internet Deployment}.
\newblock \bibinfo{howpublished}{IETF 117, Presentation to CCWG}.
\newblock


\bibitem[Casparsen et~al\mbox{.}(2026)]%
        {casparsen-leo-latency}
\bibfield{author}{\bibinfo{person}{Andreas Casparsen}, \bibinfo{person}{Jonas~Ellegaard Jakobsen}, \bibinfo{person}{Jimmy~Jessen Nielsen}, \bibinfo{person}{Petar Popovski}, {and} \bibinfo{person}{Israel Leyva-Mayorga}.} \bibinfo{year}{2026}\natexlab{}.
\newblock \showarticletitle{Statistical Characterization and Prediction of {E2E} Latency over {LEO} Satellite Networks}.
\newblock \bibinfo{journal}{\emph{npj Wireless Technology}}  \bibinfo{volume}{2} (\bibinfo{year}{2026}), \bibinfo{pages}{36}.
\newblock
\href{https://doi.org/10.1038/s44459-026-00044-z}{doi:\nolinkurl{10.1038/s44459-026-00044-z}}
\newblock
\shownote{Also arXiv:2601.08439}.


\bibitem[Cech et~al\mbox{.}(2026a)]%
        {cech-tokens}
\bibfield{author}{\bibinfo{person}{Hendrik Cech}, \bibinfo{person}{Patrick Bokelmann}, {and} \bibinfo{person}{Nitinder Mohan}.} \bibinfo{year}{2026}\natexlab{a}.
\newblock \showarticletitle{Tokens, Not Packets: Rethinking the Multipath {QUIC} Scheduling Interface}. In \bibinfo{booktitle}{\emph{Proceedings of the Applied Networking Research Workshop (ANRW)}}. \bibinfo{publisher}{ACM}, \bibinfo{pages}{140--146}.
\newblock
\href{https://doi.org/10.1145/3822163.3827925}{doi:\nolinkurl{10.1145/3822163.3827925}}


\bibitem[Cech et~al\mbox{.}(2026b)]%
        {dissectstarlink2026sigcomm}
\bibfield{author}{\bibinfo{person}{Hendrik Cech}, \bibinfo{person}{Nitinder Mohan}, {and} \bibinfo{person}{J{\"o}rg Ott}.} \bibinfo{year}{2026}\natexlab{b}.
\newblock \showarticletitle{Dissecting the {StarLink}: Characterizing Queuing and Flow Dynamics in the {Starlink} Network}. In \bibinfo{booktitle}{\emph{Proceedings of the ACM SIGCOMM 2026 Conference}}. \bibinfo{publisher}{ACM}.
\newblock
\newblock
\shownote{To appear}.


\bibitem[Chaudhary et~al\mbox{.}(2025)]%
        {compact}
\bibfield{author}{\bibinfo{person}{Shubham Chaudhary}, \bibinfo{person}{Navneet Mishra}, \bibinfo{person}{Keshav Gambhir}, \bibinfo{person}{Tanmay Rajore}, \bibinfo{person}{Arani Bhattacharya}, {and} \bibinfo{person}{Mukulika Maity}.} \bibinfo{year}{2025}\natexlab{}.
\newblock \showarticletitle{{COMPACT}: Content-aware Multipath Live Video Streaming for Online Classes using Video Tiles}. In \bibinfo{booktitle}{\emph{Proceedings of the 16th ACM Multimedia Systems Conference (MMSys)}}. \bibinfo{publisher}{ACM}, \bibinfo{pages}{201--213}.
\newblock
\href{https://doi.org/10.1145/3712676.3714451}{doi:\nolinkurl{10.1145/3712676.3714451}}


\bibitem[Clark and Tennenhouse(1990)]%
        {clark-alf}
\bibfield{author}{\bibinfo{person}{David~D. Clark} {and} \bibinfo{person}{David~L. Tennenhouse}.} \bibinfo{year}{1990}\natexlab{}.
\newblock \showarticletitle{Architectural Considerations for a New Generation of Protocols}. In \bibinfo{booktitle}{\emph{Proceedings of the ACM SIGCOMM 1990 Conference}}. \bibinfo{publisher}{ACM}, \bibinfo{pages}{200--208}.
\newblock
\href{https://doi.org/10.1145/99508.99553}{doi:\nolinkurl{10.1145/99508.99553}}


\bibitem[{Cloudflare}(2026)]%
        {cloudflare-moq-relays}
\bibfield{author}{\bibinfo{person}{{Cloudflare}}.} \bibinfo{year}{2026}\natexlab{}.
\newblock \bibinfo{title}{An {API} for {MoQ}: Provision Your Own Isolated Relays}.
\newblock \bibinfo{howpublished}{The Cloudflare Blog, \url{https://blog.cloudflare.com/moq-relays/}}.
\newblock
\newblock
\shownote{Accessed August 2026}.


\bibitem[Coninck and Bonaventure(2017)]%
        {deconinck-mpquic}
\bibfield{author}{\bibinfo{person}{Quentin~De Coninck} {and} \bibinfo{person}{Olivier Bonaventure}.} \bibinfo{year}{2017}\natexlab{}.
\newblock \showarticletitle{Multipath {QUIC}: Design and Evaluation}. In \bibinfo{booktitle}{\emph{Proceedings of the 13th International Conference on Emerging Networking Experiments and Technologies (CoNEXT '17)}}. \bibinfo{publisher}{ACM}, \bibinfo{pages}{160--166}.
\newblock
\href{https://doi.org/10.1145/3143361.3143370}{doi:\nolinkurl{10.1145/3143361.3143370}}


\bibitem[Dincer(2025)]%
        {cloudflare-moq}
\bibfield{author}{\bibinfo{person}{Renan Dincer}.} \bibinfo{year}{2025}\natexlab{}.
\newblock \bibinfo{title}{{MoQ}: Refactoring the Internet's Real-Time Media Stack}.
\newblock \bibinfo{howpublished}{The Cloudflare Blog, \url{https://blog.cloudflare.com/moq/}}.
\newblock
\newblock
\shownote{Accessed August 2026}.


\bibitem[Engelbart et~al\mbox{.}(2025)]%
        {moq-dns}
\bibfield{author}{\bibinfo{person}{Mathis Engelbart}, \bibinfo{person}{Mike Kosek}, \bibinfo{person}{Lars Eggert}, {and} \bibinfo{person}{J{\"o}rg Ott}.} \bibinfo{year}{2025}\natexlab{}.
\newblock \showarticletitle{From req/res to pub/sub: Exploring Media over {QUIC} Transport for {DNS}}. In \bibinfo{booktitle}{\emph{Proceedings of the 24th ACM Workshop on Hot Topics in Networks (HotNets)}}. \bibinfo{publisher}{ACM}, \bibinfo{pages}{419--426}.
\newblock
\href{https://doi.org/10.1145/3772356.3772416}{doi:\nolinkurl{10.1145/3772356.3772416}}


\bibitem[Ferlin et~al\mbox{.}(2016)]%
        {blest}
\bibfield{author}{\bibinfo{person}{Simone Ferlin}, \bibinfo{person}{\"{O}zg\"{u} Alay}, \bibinfo{person}{Olivier Mehani}, {and} \bibinfo{person}{Roksana Boreli}.} \bibinfo{year}{2016}\natexlab{}.
\newblock \showarticletitle{BLEST: Blocking Estimation-based MPTCP Scheduler for Heterogeneous Networks}. In \bibinfo{booktitle}{\emph{2016 IFIP Networking Conference (Networking) and Workshops}}. \bibinfo{publisher}{IEEE}, \bibinfo{pages}{431--439}.
\newblock
\href{https://doi.org/10.1109/IFIPNetworking.2016.7497206}{doi:\nolinkurl{10.1109/IFIPNetworking.2016.7497206}}


\bibitem[Fouladi et~al\mbox{.}(2018)]%
        {salsify}
\bibfield{author}{\bibinfo{person}{Sadjad Fouladi}, \bibinfo{person}{John Emmons}, \bibinfo{person}{Emre Orbay}, \bibinfo{person}{Catherine Wu}, \bibinfo{person}{Riad~S. Wahby}, {and} \bibinfo{person}{Keith Winstein}.} \bibinfo{year}{2018}\natexlab{}.
\newblock \showarticletitle{Salsify: Low-Latency Network Video through Tighter Integration between a Video Codec and a Transport Protocol}. In \bibinfo{booktitle}{\emph{15th USENIX Symposium on Networked Systems Design and Implementation (NSDI '18)}}. \bibinfo{publisher}{USENIX Association}, \bibinfo{pages}{267--282}.
\newblock


\bibitem[Freeman(2025)]%
        {moq-filtering}
\bibfield{author}{\bibinfo{person}{Andrew~C. Freeman}.} \bibinfo{year}{2025}\natexlab{}.
\newblock \showarticletitle{Toward Accessible and Safe Live Streaming Using Distributed Content Filtering with {MoQ}}. In \bibinfo{booktitle}{\emph{2025 IEEE International Conference on Multimedia and Expo Workshops (ICMEW)}}. \bibinfo{publisher}{IEEE}, \bibinfo{pages}{1--6}.
\newblock
\href{https://doi.org/10.1109/ICMEW68306.2025.11152153}{doi:\nolinkurl{10.1109/ICMEW68306.2025.11152153}}
\newblock
\shownote{LIVES Workshop}.


\bibitem[Fr{\"o}mmgen et~al\mbox{.}(2017)]%
        {progmp}
\bibfield{author}{\bibinfo{person}{Alexander Fr{\"o}mmgen}, \bibinfo{person}{Amr Rizk}, \bibinfo{person}{Tobias Erbsh{\"a}u{\ss}er}, \bibinfo{person}{Mira Weller}, \bibinfo{person}{Boris Koldehofe}, \bibinfo{person}{Alejandro Buchmann}, {and} \bibinfo{person}{Ralf Steinmetz}.} \bibinfo{year}{2017}\natexlab{}.
\newblock \showarticletitle{A Programming Model for Application-Defined Multipath {TCP} Scheduling}. In \bibinfo{booktitle}{\emph{Proceedings of the 18th ACM/IFIP/USENIX Middleware Conference}}. \bibinfo{publisher}{ACM}, \bibinfo{pages}{134--146}.
\newblock
\href{https://doi.org/10.1145/3135974.3135979}{doi:\nolinkurl{10.1145/3135974.3135979}}


\bibitem[Gottipati and Qiu(2025)]%
        {gottipati-pacing}
\bibfield{author}{\bibinfo{person}{Aashish Gottipati} {and} \bibinfo{person}{Lili Qiu}.} \bibinfo{year}{2025}\natexlab{}.
\newblock \bibinfo{title}{Dynamic Pacing for Real-time Satellite Traffic}.
\newblock \bibinfo{howpublished}{arXiv preprint}.
\newblock
\newblock
\shownote{arXiv:2507.09798}.


\bibitem[Ha et~al\mbox{.}(2008)]%
        {cubic}
\bibfield{author}{\bibinfo{person}{Sangtae Ha}, \bibinfo{person}{Injong Rhee}, {and} \bibinfo{person}{Lisong Xu}.} \bibinfo{year}{2008}\natexlab{}.
\newblock \showarticletitle{{CUBIC}: A New {TCP}-Friendly High-Speed {TCP} Variant}.
\newblock \bibinfo{journal}{\emph{ACM SIGOPS Operating Systems Review}} \bibinfo{volume}{42}, \bibinfo{number}{5} (\bibinfo{date}{July} \bibinfo{year}{2008}), \bibinfo{pages}{64--74}.
\newblock
\href{https://doi.org/10.1145/1400097.1400105}{doi:\nolinkurl{10.1145/1400097.1400105}}


\bibitem[Han et~al\mbox{.}(2016)]%
        {mp-dash}
\bibfield{author}{\bibinfo{person}{Bo Han}, \bibinfo{person}{Feng Qian}, \bibinfo{person}{Lusheng Ji}, {and} \bibinfo{person}{Vijay Gopalakrishnan}.} \bibinfo{year}{2016}\natexlab{}.
\newblock \showarticletitle{{MP-DASH}: Adaptive Video Streaming Over Preference-Aware Multipath}. In \bibinfo{booktitle}{\emph{Proceedings of the 12th International Conference on Emerging Networking EXperiments and Technologies (CoNEXT)}}. \bibinfo{publisher}{ACM}, \bibinfo{pages}{129--143}.
\newblock
\href{https://doi.org/10.1145/2999572.2999606}{doi:\nolinkurl{10.1145/2999572.2999606}}


\bibitem[Hancke(2023)]%
        {hancke-av1-meet}
\bibfield{author}{\bibinfo{person}{Philipp Hancke}.} \bibinfo{year}{2023}\natexlab{}.
\newblock \bibinfo{title}{The Hidden {AV1} Gift in {Google Meet}}.
\newblock \bibinfo{howpublished}{webrtcHacks, \url{https://webrtchacks.com/the-hidden-av1-gift-in-google-meet/}}.
\newblock
\newblock
\shownote{Accessed August 2026}.


\bibitem[{ITU-T}(2003)]%
        {itu-g114}
\bibfield{author}{\bibinfo{person}{{ITU-T}}.} \bibinfo{year}{2003}\natexlab{}.
\newblock \bibinfo{title}{One-Way Transmission Time}.
\newblock \bibinfo{howpublished}{ITU-T Recommendation G.114}.
\newblock


\bibitem[Jennings et~al\mbox{.}(2026)]%
        {moq-secure-objects}
\bibfield{author}{\bibinfo{person}{Cullen Jennings}, \bibinfo{person}{Suhas Nandakumar}, {and} \bibinfo{person}{Richard Barnes}.} \bibinfo{year}{2026}\natexlab{}.
\newblock \bibinfo{booktitle}{\emph{End-to-End Secure Objects for Media over {QUIC} Transport}}.
\newblock \bibinfo{type}{Internet-Draft} draft-ietf-moq-secure-objects-01. \bibinfo{institution}{Internet Engineering Task Force}.
\newblock
\newblock
\shownote{Work in Progress}.


\bibitem[Kozuka and Okabe(2023)]%
        {kozuka-policy}
\bibfield{author}{\bibinfo{person}{Masahiro Kozuka} {and} \bibinfo{person}{Yasuo Okabe}.} \bibinfo{year}{2023}\natexlab{}.
\newblock \showarticletitle{A Policy-Based Path Selection Mechanism in {QUIC} Multipath Extension}. In \bibinfo{booktitle}{\emph{2023 IEEE 47th Annual Computers, Software, and Applications Conference (COMPSAC)}}. \bibinfo{publisher}{IEEE}, \bibinfo{pages}{1255--1259}.
\newblock
\href{https://doi.org/10.1109/COMPSAC57700.2023.00190}{doi:\nolinkurl{10.1109/COMPSAC57700.2023.00190}}


\bibitem[Lai et~al\mbox{.}(2025)]%
        {leocc}
\bibfield{author}{\bibinfo{person}{Zeqi Lai}, \bibinfo{person}{Zonglun Li}, \bibinfo{person}{Qian Wu}, \bibinfo{person}{Hewu Li}, \bibinfo{person}{Jihao Li}, \bibinfo{person}{Xin Xie}, \bibinfo{person}{Yuanjie Li}, \bibinfo{person}{Jun Liu}, {and} \bibinfo{person}{Jianping Wu}.} \bibinfo{year}{2025}\natexlab{}.
\newblock \showarticletitle{{LeoCC}: Making Internet Congestion Control Robust to {LEO} Satellite Dynamics}. In \bibinfo{booktitle}{\emph{Proceedings of the ACM SIGCOMM 2025 Conference}}. \bibinfo{publisher}{ACM}, \bibinfo{pages}{129--146}.
\newblock
\href{https://doi.org/10.1145/3718958.3750491}{doi:\nolinkurl{10.1145/3718958.3750491}}


\bibitem[Lai et~al\mbox{.}(2024)]%
        {lai-leo-mobility}
\bibfield{author}{\bibinfo{person}{Zeqi Lai}, \bibinfo{person}{Zonglun Li}, \bibinfo{person}{Qian Wu}, \bibinfo{person}{Hewu Li}, \bibinfo{person}{Weisen Liu}, \bibinfo{person}{Yijie Liu}, \bibinfo{person}{Xin Xie}, \bibinfo{person}{Yuanjie Li}, {and} \bibinfo{person}{Jun Liu}.} \bibinfo{year}{2024}\natexlab{}.
\newblock \showarticletitle{Mind the Misleading Effects of {LEO} Mobility on End-to-End Congestion Control}. In \bibinfo{booktitle}{\emph{Proceedings of the 23rd ACM Workshop on Hot Topics in Networks (HotNets)}}. \bibinfo{publisher}{ACM}, \bibinfo{pages}{34--42}.
\newblock
\href{https://doi.org/10.1145/3696348.3696867}{doi:\nolinkurl{10.1145/3696348.3696867}}


\bibitem[Lee et~al\mbox{.}(2026)]%
        {qcon}
\bibfield{author}{\bibinfo{person}{Goodsol Lee}, \bibinfo{person}{Junhong Min}, \bibinfo{person}{Seyeon Kim}, \bibinfo{person}{Juheon Yi}, \bibinfo{person}{Kwang~Taik Kim}, \bibinfo{person}{Mung Chiang}, \bibinfo{person}{Sangtae Ha}, \bibinfo{person}{Kyunghan Lee}, {and} \bibinfo{person}{Saewoong Bahk}.} \bibinfo{year}{2026}\natexlab{}.
\newblock \showarticletitle{{QCON}: Seamless {QoE}-Aware {5G} Streaming via Multi-Connectivity}. In \bibinfo{booktitle}{\emph{Proceedings of the 23rd USENIX Symposium on Networked Systems Design and Implementation (NSDI)}}. \bibinfo{publisher}{USENIX Association}.
\newblock


\bibitem[Lin et~al\mbox{.}(2022)]%
        {gso-simulcast}
\bibfield{author}{\bibinfo{person}{Xianshang Lin}, \bibinfo{person}{Yunfei Ma}, \bibinfo{person}{Junshao Zhang}, \bibinfo{person}{Yao Cui}, \bibinfo{person}{Jing Li}, \bibinfo{person}{Shi Bai}, \bibinfo{person}{Ziyue Zhang}, \bibinfo{person}{Dennis Cai}, \bibinfo{person}{Hongqiang~Harry Liu}, {and} \bibinfo{person}{Ming Zhang}.} \bibinfo{year}{2022}\natexlab{}.
\newblock \showarticletitle{{GSO-Simulcast}: Global Stream Orchestration in Simulcast Video Conferencing Systems}. In \bibinfo{booktitle}{\emph{Proceedings of the ACM SIGCOMM 2022 Conference}}. \bibinfo{publisher}{ACM}, \bibinfo{pages}{826--839}.
\newblock
\href{https://doi.org/10.1145/3544216.3544228}{doi:\nolinkurl{10.1145/3544216.3544228}}


\bibitem[Liu et~al\mbox{.}(2026)]%
        {mpquic-draft}
\bibfield{author}{\bibinfo{person}{Yanmei Liu}, \bibinfo{person}{Yunfei Ma}, \bibinfo{person}{Quentin~De Coninck}, \bibinfo{person}{Olivier Bonaventure}, \bibinfo{person}{Christian Huitema}, {and} \bibinfo{person}{Mirja K\"uhlewind}.} \bibinfo{year}{2026}\natexlab{}.
\newblock \bibinfo{booktitle}{\emph{Managing Multiple Paths for a {QUIC} Connection}}.
\newblock \bibinfo{type}{Internet-Draft} draft-ietf-quic-multipath-21. \bibinfo{institution}{Internet Engineering Task Force}.
\newblock
\newblock
\shownote{Work in Progress, RFC Editor Queue}.


\bibitem[McKeown et~al\mbox{.}(2008)]%
        {mckeown-openflow}
\bibfield{author}{\bibinfo{person}{Nick McKeown}, \bibinfo{person}{Tom Anderson}, \bibinfo{person}{Hari Balakrishnan}, \bibinfo{person}{Guru Parulkar}, \bibinfo{person}{Larry Peterson}, \bibinfo{person}{Jennifer Rexford}, \bibinfo{person}{Scott Shenker}, {and} \bibinfo{person}{Jonathan Turner}.} \bibinfo{year}{2008}\natexlab{}.
\newblock \showarticletitle{{OpenFlow}: Enabling Innovation in Campus Networks}.
\newblock \bibinfo{journal}{\emph{ACM SIGCOMM Computer Communication Review}} \bibinfo{volume}{38}, \bibinfo{number}{2} (\bibinfo{date}{April} \bibinfo{year}{2008}), \bibinfo{pages}{69--74}.
\newblock
\href{https://doi.org/10.1145/1355734.1355746}{doi:\nolinkurl{10.1145/1355734.1355746}}


\bibitem[Meng et~al\mbox{.}(2022)]%
        {zhuge}
\bibfield{author}{\bibinfo{person}{Zili Meng}, \bibinfo{person}{Yaning Guo}, \bibinfo{person}{Chen Sun}, \bibinfo{person}{Bo Wang}, \bibinfo{person}{Justine Sherry}, \bibinfo{person}{Hongqiang~Harry Liu}, {and} \bibinfo{person}{Mingwei Xu}.} \bibinfo{year}{2022}\natexlab{}.
\newblock \showarticletitle{Achieving Consistent Low Latency for Wireless Real-Time Communications with the Shortest Control Loop}. In \bibinfo{booktitle}{\emph{Proceedings of the ACM SIGCOMM 2022 Conference}}. \bibinfo{publisher}{ACM}, \bibinfo{pages}{193--206}.
\newblock
\href{https://doi.org/10.1145/3544216.3544225}{doi:\nolinkurl{10.1145/3544216.3544225}}


\bibitem[Michel et~al\mbox{.}(2025)]%
        {scallop}
\bibfield{author}{\bibinfo{person}{Oliver Michel}, \bibinfo{person}{Satadal Sengupta}, \bibinfo{person}{Hyojoon Kim}, \bibinfo{person}{Ravi Netravali}, {and} \bibinfo{person}{Jennifer Rexford}.} \bibinfo{year}{2025}\natexlab{}.
\newblock \showarticletitle{Scalable Video Conferencing Using {SDN} Principles}. In \bibinfo{booktitle}{\emph{Proceedings of the ACM SIGCOMM 2025 Conference}}. \bibinfo{publisher}{ACM}, \bibinfo{pages}{1213--1231}.
\newblock
\href{https://doi.org/10.1145/3718958.3750489}{doi:\nolinkurl{10.1145/3718958.3750489}}


\bibitem[Mohan et~al\mbox{.}(2024)]%
        {mohan-starlink}
\bibfield{author}{\bibinfo{person}{Nitinder Mohan}, \bibinfo{person}{Andrew~E. Ferguson}, \bibinfo{person}{Hendrik Cech}, \bibinfo{person}{Rohan Bose}, \bibinfo{person}{Prakita~Rayyan Renatin}, \bibinfo{person}{Mahesh~K. Marina}, {and} \bibinfo{person}{J\"{o}rg Ott}.} \bibinfo{year}{2024}\natexlab{}.
\newblock \showarticletitle{A Multifaceted Look at {Starlink} Performance}. In \bibinfo{booktitle}{\emph{Proceedings of the ACM Web Conference 2024 (WWW '24)}}. \bibinfo{publisher}{ACM}, \bibinfo{pages}{2723--2734}.
\newblock
\href{https://doi.org/10.1145/3589334.3645328}{doi:\nolinkurl{10.1145/3589334.3645328}}


\bibitem[Nandakumar et~al\mbox{.}(2026)]%
        {moq-transport}
\bibfield{author}{\bibinfo{person}{Suhas Nandakumar}, \bibinfo{person}{Victor Vasiliev}, \bibinfo{person}{Ian Swett}, {and} \bibinfo{person}{Alan Frindell}.} \bibinfo{year}{2026}\natexlab{}.
\newblock \bibinfo{booktitle}{\emph{Media over QUIC Transport}}.
\newblock \bibinfo{type}{Internet-Draft} draft-ietf-moq-transport-19. \bibinfo{institution}{Internet Engineering Task Force}.
\newblock
\newblock
\shownote{Work in Progress}.


\bibitem[Narayanan et~al\mbox{.}(2021)]%
        {narayanan-5g-variegated}
\bibfield{author}{\bibinfo{person}{Arvind Narayanan}, \bibinfo{person}{Xumiao Zhang}, \bibinfo{person}{Ruiyang Zhu}, \bibinfo{person}{Ahmad Hassan}, \bibinfo{person}{Shuowei Jin}, \bibinfo{person}{Xiao Zhu}, \bibinfo{person}{Xiaoxuan Zhang}, \bibinfo{person}{Denis Rybkin}, \bibinfo{person}{Zhengxuan Yang}, \bibinfo{person}{Zhuoqing~Morley Mao}, \bibinfo{person}{Feng Qian}, {and} \bibinfo{person}{Zhi-Li Zhang}.} \bibinfo{year}{2021}\natexlab{}.
\newblock \showarticletitle{A Variegated Look at {5G} in the Wild: Performance, Power, and {QoE} Implications}. In \bibinfo{booktitle}{\emph{Proceedings of the ACM SIGCOMM 2021 Conference (SIGCOMM '21)}}. \bibinfo{publisher}{ACM}, \bibinfo{pages}{610--625}.
\newblock
\href{https://doi.org/10.1145/3452296.3472923}{doi:\nolinkurl{10.1145/3452296.3472923}}


\bibitem[N{\'e}meth et~al\mbox{.}(2025)]%
        {nemeth-failover}
\bibfield{author}{\bibinfo{person}{Felici{\'a}n N{\'e}meth}, \bibinfo{person}{Zolt{\'a}n Szatm{\'a}ry}, \bibinfo{person}{Istv{\'a}n Pelle}, {and} \bibinfo{person}{Tam{\'a}s L{\'e}vai}.} \bibinfo{year}{2025}\natexlab{}.
\newblock \showarticletitle{{MoQ} Resilience: Implicit Fast Failover}. In \bibinfo{booktitle}{\emph{Proceedings of the 3rd Workshop on Emerging Multimedia Systems (EMS)}}. \bibinfo{publisher}{ACM}, \bibinfo{pages}{61--63}.
\newblock
\href{https://doi.org/10.1145/3746441.3748235}{doi:\nolinkurl{10.1145/3746441.3748235}}


\bibitem[Nygren et~al\mbox{.}(2010)]%
        {nygren-akamai}
\bibfield{author}{\bibinfo{person}{Erik Nygren}, \bibinfo{person}{Ramesh~K. Sitaraman}, {and} \bibinfo{person}{Jennifer Sun}.} \bibinfo{year}{2010}\natexlab{}.
\newblock \showarticletitle{The {Akamai} Network: A Platform for High-Performance Internet Applications}.
\newblock \bibinfo{journal}{\emph{ACM SIGOPS Operating Systems Review}} \bibinfo{volume}{44}, \bibinfo{number}{3} (\bibinfo{year}{2010}), \bibinfo{pages}{2--19}.
\newblock
\href{https://doi.org/10.1145/1842733.1842736}{doi:\nolinkurl{10.1145/1842733.1842736}}


\bibitem[{OpenMOQ Software Consortium}(2026)]%
        {openmoq}
\bibfield{author}{\bibinfo{person}{{OpenMOQ Software Consortium}}.} \bibinfo{year}{2026}\natexlab{}.
\newblock \bibinfo{title}{{OpenMOQ} Software Consortium}.
\newblock \bibinfo{howpublished}{\url{https://openmoq.org/}}.
\newblock
\newblock
\shownote{Accessed August 2026}.


\bibitem[Paasch et~al\mbox{.}(2014)]%
        {paasch-schedulers}
\bibfield{author}{\bibinfo{person}{Christoph Paasch}, \bibinfo{person}{Simone Ferlin}, \bibinfo{person}{\"{O}zg\"{u} Alay}, {and} \bibinfo{person}{Olivier Bonaventure}.} \bibinfo{year}{2014}\natexlab{}.
\newblock \showarticletitle{Experimental Evaluation of Multipath {TCP} Schedulers}. In \bibinfo{booktitle}{\emph{Proceedings of the 2014 ACM SIGCOMM Workshop on Capacity Sharing (CSWS '14)}}. \bibinfo{publisher}{ACM}, \bibinfo{pages}{27--32}.
\newblock
\href{https://doi.org/10.1145/2630088.2631977}{doi:\nolinkurl{10.1145/2630088.2631977}}


\bibitem[Pan et~al\mbox{.}(2026)]%
        {syntra}
\bibfield{author}{\bibinfo{person}{Jia Pan}, \bibinfo{person}{Anup Agarwal}, \bibinfo{person}{I\c{s}\i l Dillig}, {and} \bibinfo{person}{Venkat Arun}.} \bibinfo{year}{2026}\natexlab{}.
\newblock \showarticletitle{{Syntra}: Synthesizing Cross-Layer Controllers for Low-Latency Video Streaming}. In \bibinfo{booktitle}{\emph{Proceedings of the 23rd USENIX Symposium on Networked Systems Design and Implementation (NSDI)}}. \bibinfo{publisher}{USENIX Association}.
\newblock


\bibitem[Pie{\v s}ka et~al\mbox{.}(2024)]%
        {pieska-nested}
\bibfield{author}{\bibinfo{person}{Marcus Pie{\v s}ka}, \bibinfo{person}{Andreas Kassler}, \bibinfo{person}{Anna Brunstrom}, \bibinfo{person}{Veselin Rako{\v c}evi{\'c}}, {and} \bibinfo{person}{Markus Amend}.} \bibinfo{year}{2024}\natexlab{}.
\newblock \showarticletitle{Performance Impact of Nested Congestion Control on Transport-Layer Multipath Tunneling}.
\newblock \bibinfo{journal}{\emph{Future Internet}} \bibinfo{volume}{16}, \bibinfo{number}{7} (\bibinfo{year}{2024}), \bibinfo{pages}{233}.
\newblock
\href{https://doi.org/10.3390/fi16070233}{doi:\nolinkurl{10.3390/fi16070233}}


\bibitem[Rabitsch et~al\mbox{.}(2018)]%
        {rabitsch-stream}
\bibfield{author}{\bibinfo{person}{Alexander Rabitsch}, \bibinfo{person}{Per Hurtig}, {and} \bibinfo{person}{Anna Brunstrom}.} \bibinfo{year}{2018}\natexlab{}.
\newblock \showarticletitle{A Stream-Aware Multipath {QUIC} Scheduler for Heterogeneous Paths}. In \bibinfo{booktitle}{\emph{Proceedings of the Workshop on the Evolution, Performance, and Interoperability of QUIC (EPIQ)}}. \bibinfo{publisher}{ACM}, \bibinfo{pages}{29--35}.
\newblock
\href{https://doi.org/10.1145/3284850.3284855}{doi:\nolinkurl{10.1145/3284850.3284855}}


\bibitem[Raiciu et~al\mbox{.}(2012)]%
        {raiciu-mptcp}
\bibfield{author}{\bibinfo{person}{Costin Raiciu}, \bibinfo{person}{Christoph Paasch}, \bibinfo{person}{Sebastien Barre}, \bibinfo{person}{Alan Ford}, \bibinfo{person}{Michio Honda}, \bibinfo{person}{Fabien Duchene}, \bibinfo{person}{Olivier Bonaventure}, {and} \bibinfo{person}{Mark Handley}.} \bibinfo{year}{2012}\natexlab{}.
\newblock \showarticletitle{How Hard Can It Be? Designing and Implementing a Deployable Multipath {TCP}}. In \bibinfo{booktitle}{\emph{9th USENIX Symposium on Networked Systems Design and Implementation (NSDI '12)}}. \bibinfo{publisher}{USENIX Association}, \bibinfo{pages}{399--412}.
\newblock


\bibitem[Ram{\'i}rez-Arroyo et~al\mbox{.}(2026)]%
        {ramirez-multiconnectivity}
\bibfield{author}{\bibinfo{person}{Alejandro Ram{\'i}rez-Arroyo}, \bibinfo{person}{O.~S. Pe{\~n}aherrera-Pulla}, {and} \bibinfo{person}{Preben Mogensen}.} \bibinfo{year}{2026}\natexlab{}.
\newblock \showarticletitle{Toward Reliable Connectivity: Measurement-Driven Assessment of Starlink and OneWeb Non-Terrestrial and {5G} Terrestrial Networks}.
\newblock \bibinfo{journal}{\emph{IEEE Open Journal of the Communications Society}}  \bibinfo{volume}{7} (\bibinfo{year}{2026}), \bibinfo{pages}{3956--3973}.
\newblock
\href{https://doi.org/10.1109/OJCOMS.2026.3682548}{doi:\nolinkurl{10.1109/OJCOMS.2026.3682548}}


\bibitem[Saltzer et~al\mbox{.}(1984)]%
        {saltzer-e2e}
\bibfield{author}{\bibinfo{person}{Jerome~H. Saltzer}, \bibinfo{person}{David~P. Reed}, {and} \bibinfo{person}{David~D. Clark}.} \bibinfo{year}{1984}\natexlab{}.
\newblock \showarticletitle{End-to-End Arguments in System Design}.
\newblock \bibinfo{journal}{\emph{ACM Transactions on Computer Systems}} \bibinfo{volume}{2}, \bibinfo{number}{4} (\bibinfo{date}{Nov.} \bibinfo{year}{1984}), \bibinfo{pages}{277--288}.
\newblock
\href{https://doi.org/10.1145/357401.357402}{doi:\nolinkurl{10.1145/357401.357402}}


\bibitem[Sathyanarayana et~al\mbox{.}(2023)]%
        {converge}
\bibfield{author}{\bibinfo{person}{Sandesh~Dhawaskar Sathyanarayana}, \bibinfo{person}{Kyunghan Lee}, \bibinfo{person}{Dirk Grunwald}, {and} \bibinfo{person}{Sangtae Ha}.} \bibinfo{year}{2023}\natexlab{}.
\newblock \showarticletitle{Converge: {QoE}-driven Multipath Video Conferencing over {WebRTC}}. In \bibinfo{booktitle}{\emph{Proceedings of the ACM SIGCOMM 2023 Conference}}. \bibinfo{publisher}{ACM}, \bibinfo{pages}{637--653}.
\newblock
\href{https://doi.org/10.1145/3603269.3604822}{doi:\nolinkurl{10.1145/3603269.3604822}}


\bibitem[Scharf and Ford(2013)]%
        {rfc6897}
\bibfield{author}{\bibinfo{person}{Michael Scharf} {and} \bibinfo{person}{Alan Ford}.} \bibinfo{year}{2013}\natexlab{}.
\newblock \bibinfo{booktitle}{\emph{Multipath {TCP} ({MPTCP}) Application Interface Considerations}}.
\newblock \bibinfo{type}{RFC} 6897. \bibinfo{institution}{Internet Engineering Task Force}.
\newblock
\href{https://doi.org/10.17487/RFC6897}{doi:\nolinkurl{10.17487/RFC6897}}


\bibitem[Schwarz et~al\mbox{.}(2007)]%
        {svc-overview}
\bibfield{author}{\bibinfo{person}{Heiko Schwarz}, \bibinfo{person}{Detlev Marpe}, {and} \bibinfo{person}{Thomas Wiegand}.} \bibinfo{year}{2007}\natexlab{}.
\newblock \showarticletitle{Overview of the Scalable Video Coding Extension of the H.264/AVC Standard}.
\newblock \bibinfo{journal}{\emph{IEEE Transactions on Circuits and Systems for Video Technology}} \bibinfo{volume}{17}, \bibinfo{number}{9} (\bibinfo{year}{2007}), \bibinfo{pages}{1103--1120}.
\newblock
\href{https://doi.org/10.1109/TCSVT.2007.905532}{doi:\nolinkurl{10.1109/TCSVT.2007.905532}}


\bibitem[Shi et~al\mbox{.}(2019)]%
        {dtp}
\bibfield{author}{\bibinfo{person}{Hang Shi}, \bibinfo{person}{Yong Cui}, \bibinfo{person}{Feng Qian}, {and} \bibinfo{person}{Yuming Hu}.} \bibinfo{year}{2019}\natexlab{}.
\newblock \showarticletitle{{DTP}: Deadline-aware Transport Protocol}. In \bibinfo{booktitle}{\emph{Proceedings of the 3rd Asia-Pacific Workshop on Networking (APNet)}}. \bibinfo{publisher}{ACM}, \bibinfo{pages}{1--7}.
\newblock
\href{https://doi.org/10.1145/3343180.3343191}{doi:\nolinkurl{10.1145/3343180.3343191}}


\bibitem[Shi et~al\mbox{.}(2020)]%
        {pstream}
\bibfield{author}{\bibinfo{person}{Xiang Shi}, \bibinfo{person}{Lin Wang}, \bibinfo{person}{Fa Zhang}, \bibinfo{person}{Biyu Zhou}, {and} \bibinfo{person}{Zhiyong Liu}.} \bibinfo{year}{2020}\natexlab{}.
\newblock \showarticletitle{{PStream}: Priority-Based Stream Scheduling for Heterogeneous Paths in Multipath-{QUIC}}. In \bibinfo{booktitle}{\emph{2020 29th International Conference on Computer Communications and Networks (ICCCN)}}. \bibinfo{publisher}{IEEE}, \bibinfo{pages}{1--8}.
\newblock
\href{https://doi.org/10.1109/ICCCN49398.2020.9209682}{doi:\nolinkurl{10.1109/ICCCN49398.2020.9209682}}


\bibitem[Shreedhar et~al\mbox{.}(2018)]%
        {shreedhar2018qaware}
\bibfield{author}{\bibinfo{person}{Tanya Shreedhar}, \bibinfo{person}{Nitinder Mohan}, \bibinfo{person}{Sanjit~K. Kaul}, {and} \bibinfo{person}{Jussi Kangasharju}.} \bibinfo{year}{2018}\natexlab{}.
\newblock \showarticletitle{{QAware}: A Cross-Layer Approach to {MPTCP} Scheduling}. In \bibinfo{booktitle}{\emph{2018 IFIP Networking Conference (IFIP Networking) and Workshops}}. \bibinfo{publisher}{IEEE}, \bibinfo{pages}{1--9}.
\newblock
\href{https://doi.org/10.23919/IFIPNetworking.2018.8696843}{doi:\nolinkurl{10.23919/IFIPNetworking.2018.8696843}}


\bibitem[Singh et~al\mbox{.}(2013)]%
        {mprtp}
\bibfield{author}{\bibinfo{person}{Varun Singh}, \bibinfo{person}{Saba Ahsan}, {and} \bibinfo{person}{J{\"o}rg Ott}.} \bibinfo{year}{2013}\natexlab{}.
\newblock \showarticletitle{{MPRTP}: Multipath Considerations for Real-time Media}. In \bibinfo{booktitle}{\emph{Proceedings of the 4th ACM Multimedia Systems Conference (MMSys)}}. \bibinfo{publisher}{ACM}, \bibinfo{pages}{190--201}.
\newblock
\href{https://doi.org/10.1145/2483977.2484002}{doi:\nolinkurl{10.1145/2483977.2484002}}


\bibitem[Song et~al\mbox{.}(2026)]%
        {camp-multipath}
\bibfield{author}{\bibinfo{person}{Yuhong Song}, \bibinfo{person}{Changlong Li}, {and} \bibinfo{person}{Qing Li}.} \bibinfo{year}{2026}\natexlab{}.
\newblock \bibinfo{booktitle}{\emph{Consistency-Aware Multipath Transport ({CAMP}) toward Interactive Multimodal {LLM}-Based Systems}}.
\newblock \bibinfo{type}{Internet-Draft} draft-song-tsvwg-camp-00. \bibinfo{institution}{Internet Engineering Task Force}.
\newblock
\newblock
\shownote{Work in Progress}.


\bibitem[sup Lim et~al\mbox{.}(2017)]%
        {ecf}
\bibfield{author}{\bibinfo{person}{Yeon sup Lim}, \bibinfo{person}{Erich~M. Nahum}, \bibinfo{person}{Don Towsley}, {and} \bibinfo{person}{Richard~J. Gibbens}.} \bibinfo{year}{2017}\natexlab{}.
\newblock \showarticletitle{{ECF}: An {MPTCP} Path Scheduler to Manage Heterogeneous Paths}. In \bibinfo{booktitle}{\emph{Proceedings of the 13th International Conference on Emerging Networking Experiments and Technologies (CoNEXT '17)}}. \bibinfo{publisher}{ACM}, \bibinfo{pages}{147--159}.
\newblock
\href{https://doi.org/10.1145/3143361.3143376}{doi:\nolinkurl{10.1145/3143361.3143376}}


\bibitem[Tanveer et~al\mbox{.}(2023)]%
        {tanveer-constellations}
\bibfield{author}{\bibinfo{person}{Hammas~Bin Tanveer}, \bibinfo{person}{Mike Puchol}, \bibinfo{person}{Rachee Singh}, \bibinfo{person}{Antonio Bianchi}, {and} \bibinfo{person}{Rishab Nithyanand}.} \bibinfo{year}{2023}\natexlab{}.
\newblock \showarticletitle{Making Sense of Constellations: Methodologies for Understanding {Starlink}'s Scheduling Algorithms}. In \bibinfo{booktitle}{\emph{Proceedings of the 2023 ACM Conference on Emerging Networking Experiments and Technologies (CoNEXT '23) Companion}}. \bibinfo{publisher}{ACM}, \bibinfo{pages}{37--43}.
\newblock
\href{https://doi.org/10.1145/3624354.3630586}{doi:\nolinkurl{10.1145/3624354.3630586}}


\bibitem[{Tencent}(2026)]%
        {tquic}
\bibfield{author}{\bibinfo{person}{{Tencent}}.} \bibinfo{year}{2026}\natexlab{}.
\newblock \bibinfo{title}{{TQUIC}: A High-Performance, Lightweight, and Cross-Platform Library for the {IETF} {QUIC} Protocol}.
\newblock \bibinfo{howpublished}{\url{https://github.com/Tencent/tquic}}.
\newblock
\newblock
\shownote{Accessed: 2026-08-07}.


\bibitem[Wang et~al\mbox{.}(2026)]%
        {alcs}
\bibfield{author}{\bibinfo{person}{Lin Wang}, \bibinfo{person}{Ze Wang}, \bibinfo{person}{Zeyi Deng}, \bibinfo{person}{Jingjing Zhang}, {and} \bibinfo{person}{Yue Gao}.} \bibinfo{year}{2026}\natexlab{}.
\newblock \showarticletitle{{ALCS}: An Adaptive Latency Compensation Scheduler for Multipath {TCP} in Satellite-Terrestrial Integrated Networks}.
\newblock \bibinfo{journal}{\emph{IEEE Transactions on Mobile Computing}} \bibinfo{volume}{25}, \bibinfo{number}{1} (\bibinfo{year}{2026}), \bibinfo{pages}{660--673}.
\newblock
\href{https://doi.org/10.1109/TMC.2025.3594896}{doi:\nolinkurl{10.1109/TMC.2025.3594896}}


\bibitem[Wiegand et~al\mbox{.}(2003)]%
        {wiegand-avc}
\bibfield{author}{\bibinfo{person}{Thomas Wiegand}, \bibinfo{person}{Gary~J. Sullivan}, \bibinfo{person}{Gisle Bjontegaard}, {and} \bibinfo{person}{Ajay Luthra}.} \bibinfo{year}{2003}\natexlab{}.
\newblock \showarticletitle{Overview of the {H.264/AVC} Video Coding Standard}.
\newblock \bibinfo{journal}{\emph{IEEE Transactions on Circuits and Systems for Video Technology}} \bibinfo{volume}{13}, \bibinfo{number}{7} (\bibinfo{date}{July} \bibinfo{year}{2003}), \bibinfo{pages}{560--576}.
\newblock
\href{https://doi.org/10.1109/TCSVT.2003.815165}{doi:\nolinkurl{10.1109/TCSVT.2003.815165}}


\bibitem[Zanaty et~al\mbox{.}(2025)]%
        {rfc9626}
\bibfield{author}{\bibinfo{person}{Mo Zanaty}, \bibinfo{person}{Espen Berger}, {and} \bibinfo{person}{Suhas Nandakumar}.} \bibinfo{year}{2025}\natexlab{}.
\newblock \bibinfo{booktitle}{\emph{Video Frame Marking {RTP} Header Extension}}.
\newblock \bibinfo{type}{RFC} 9626. \bibinfo{institution}{Internet Engineering Task Force}.
\newblock
\href{https://doi.org/10.17487/RFC9626}{doi:\nolinkurl{10.17487/RFC9626}}


\bibitem[Zheng et~al\mbox{.}(2021)]%
        {xlink}
\bibfield{author}{\bibinfo{person}{Zhilong Zheng}, \bibinfo{person}{Yunfei Ma}, \bibinfo{person}{Yanmei Liu}, \bibinfo{person}{Furong Yang}, \bibinfo{person}{Zhenyu Li}, \bibinfo{person}{Yuanbo Zhang}, \bibinfo{person}{Jiuhai Zhang}, \bibinfo{person}{Wei Shi}, \bibinfo{person}{Wentao Chen}, \bibinfo{person}{Ding Li}, \bibinfo{person}{Qing An}, \bibinfo{person}{Hai Hong}, \bibinfo{person}{Hongqiang~Harry Liu}, {and} \bibinfo{person}{Ming Zhang}.} \bibinfo{year}{2021}\natexlab{}.
\newblock \showarticletitle{{XLINK}: {QoE}-Driven Multi-Path {QUIC} Transport in Large-scale Video Services}. In \bibinfo{booktitle}{\emph{Proceedings of the ACM SIGCOMM 2021 Conference}}. \bibinfo{publisher}{ACM}, \bibinfo{pages}{418--432}.
\newblock
\href{https://doi.org/10.1145/3452296.3472893}{doi:\nolinkurl{10.1145/3452296.3472893}}


\bibitem[Zhou et~al\mbox{.}(2024)]%
        {augur}
\bibfield{author}{\bibinfo{person}{Yuhan Zhou}, \bibinfo{person}{Tingfeng Wang}, \bibinfo{person}{Liying Wang}, \bibinfo{person}{Nian Wen}, \bibinfo{person}{Rui Han}, \bibinfo{person}{Jing Wang}, \bibinfo{person}{Chenglei Wu}, \bibinfo{person}{Jiafeng Chen}, \bibinfo{person}{Longwei Jiang}, \bibinfo{person}{Shibo Wang}, \bibinfo{person}{Honghao Liu}, {and} \bibinfo{person}{Chenren Xu}.} \bibinfo{year}{2024}\natexlab{}.
\newblock \showarticletitle{{AUGUR}: Practical Mobile Multipath Transport Service for Low Tail Latency in Real-Time Streaming}. In \bibinfo{booktitle}{\emph{Proceedings of the 21st USENIX Symposium on Networked Systems Design and Implementation (NSDI)}}. \bibinfo{publisher}{USENIX Association}, \bibinfo{pages}{1901--1916}.
\newblock


\end{thebibliography}

\appendix
\makeatletter
\renewenvironment{figure}[1][]{\@float{figure}[!htb]}{\end@float}
\renewenvironment{table}[1][]{\@float{table}[!htb]}{\end@float}
\makeatother
\section{Supplementary Material}
\label{sec:appendix}

\subsection{Encoding Parameters}
\label{app:encoding}

\Cref{tab:svc_frame_sizes} gives the per-frame-type breakdown behind the size asymmetry of
\cref{fig:momq_challenges}, including the packet count each frame type occupies at the QUIC
maximum datagram size used throughout the evaluation.
An \iframe occupies 161~datagrams while no \pframe needs more than 11, so an \iframe spans several congestion-window rounds on either access link and every \pframe fits within one.

\begin{table}[t]
\centering
\caption{Typical frame sizes for 1080p SVC at 50~fps with QP-based quality differentiation (QP 18/26/42 for Layer 0/1/2).}
\label{tab:svc_frame_sizes}
\small
\begin{tabularx}{\columnwidth}{@{}L c c c@{}}
\toprule
\textbf{Frame type} & \textbf{Size} & \textbf{Packets (1430\,B)} & \textbf{Frequency} \\
\midrule
\iframe (IDR) & $\approx$230~KB & 161 & 1 per GOP \\
Layer 0 \pframe & $\approx$15~KB & 11 & 12 per GOP \\
Layer 1 \pframe & $\approx$10~KB & 7 & 12 per GOP \\
Layer 2 \pframe & $\approx$8~KB & 6 & 25 per GOP \\
\bottomrule
\end{tabularx}
\end{table}

\subsection{Object Format and Wire Elements}
\label{app:format}

\begin{figure}[t]
\centering
\includegraphics[width=\columnwidth]{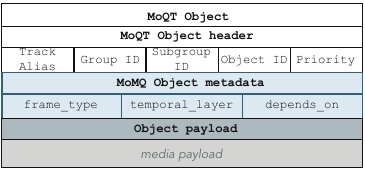}
\caption{Layout of a \moqt Object carrying \momq metadata, with the metadata block between the unchanged Object header and the payload.}
\label{fig:momq_object_format}
\end{figure}

\Cref{fig:momq_object_format} shows where the metadata that rules match against sits inside a \moqt Object.
The \moqt Object header is untouched, carrying the Track Alias, Group ID, Subgroup ID, Object ID,
and publisher priority that the base protocol already defines.
\momq adds one block between that header and the payload, holding the key-value pairs the
publisher attaches when it creates the Object, such as \texttt{frame\_type},
\texttt{temporal\_layer}, and \texttt{depends\_on}.
Keys are UTF-8 strings and values are compared as opaque byte strings, so the relay reads this
block for matching and never descends into the payload beneath it.
Since matching never touches the payload, the media may stay encrypted end to end.
Four such pairs add roughly 40--80~bytes to an Object whose payload spans 5--230~KB.

\begin{table}[t]
\centering
\caption{The five wire-level elements \momq adds to \moqt.}
\label{tab:wire_elements}
\small
\begin{tabularx}{\columnwidth}{@{}l l L@{}}
\toprule
\textbf{Element} & \textbf{Code} & \textbf{Purpose} \\
\midrule
\texttt{ENABLE\_MOMQ} & \texttt{0x10} & Setup parameter sent by both endpoints to activate the extension \\
\texttt{PATH\_MAPPING\_RULE} & \texttt{0x50} & Subscriber installs or removes a rule by identifier \\
\texttt{PATH\_MAPPING\_RESULT} & \texttt{0x51} & Relay accepts or rejects a rule operation \\
\texttt{PATH\_STATE\_REPORT} & \texttt{0x52} & Relay reports its paths, their status, and their labels \\
\texttt{PATH\_LABEL\_UPDATE} & \texttt{0x53} & Subscriber annotates a path with local knowledge \\
\bottomrule
\end{tabularx}
\end{table}

\Cref{tab:wire_elements} lists the five elements of \cref{subsec:protocol_surface} and their code points.
Each message is a flat record of counted lists, and every field is a variable-length integer, a fixed-width code, or an opaque byte string, so a relay parses all five elements without interpreting a single application-supplied value.
\texttt{ENABLE\_MOMQ} is negotiated independently of multipath, so a subscriber can install rules while the session still runs on a single path, and the preferences take effect without a further round trip once a second path is added.
A \momq control message on a session that never negotiated the parameter is a protocol violation that closes the session.
A \texttt{PATH\_MAPPING\_RULE} carries a rule identifier, an operation that is either
\texttt{INSTALL} or \texttt{REMOVE}, a counted list of Match Entries, and a counted list of Action
Entries.
\texttt{INSTALL} creates a rule or replaces the rule with the same identifier, \texttt{REMOVE} deletes it, and the relay discards every rule when the session ends.
A Match Entry is a key, one of the two operators of \cref{tab:match_operators}, and a value, while
an Action Entry is one of the four types of \cref{tab:action_types} followed by a length and its
parameters.
Two operators keep the match language small enough that every relay supports the full matching capability at bounded forwarding-path cost.
The length field lets a relay skip an action type it does not implement.

\begin{table}[t]
\centering
\caption{The two match operators.}
\label{tab:match_operators}
\small
\begin{tabularx}{\columnwidth}{@{}l l L@{}}
\toprule
\textbf{Value} & \textbf{Name} & \textbf{Semantics} \\
\midrule
0x00 & EQUALS & Exact byte-for-byte match between the metadata value and the specified value \\
0x01 & EXISTS & The metadata key is present in the Object \\
\bottomrule
\end{tabularx}
\end{table}

\begin{table}[t]
\centering
\caption{Action types for expressing delivery preferences.}
\label{tab:action_types}
\small
\begin{tabularx}{\columnwidth}{@{}l l L@{}}
\toprule
\textbf{Type} & \textbf{Name} & \textbf{Description} \\
\midrule
0x01 & PRIORITY & Scheduling urgency, with higher values more urgent \\
0x02 & BALANCING & SINGLE\_PATH or MULTI\_PATH distribution \\
0x03 & PATH\_PREFERENCE & Prefer paths whose labels match a key-value pair \\
0x04 & PATH\_AFFINITY & Co-locate with a previously forwarded Object \\
\bottomrule
\end{tabularx}
\end{table}
A \texttt{PATH\_STATE\_REPORT} carries a sequence number and a counted list of path entries, each
holding a path identifier, a status that is \texttt{ACTIVE}, \texttt{DEGRADED}, or
\texttt{UNAVAILABLE}, and that path's labels.
The sequence number is monotonic, so a subscriber drops any report older than the last one it processed.
A \texttt{PATH\_LABEL\_UPDATE} carries a path identifier and a counted list of labels, each a key
and a value.
A subscriber-assigned label overrides a relay-assigned label with the same key, and subsequent reports echo it back.
A \texttt{PATH\_MAPPING\_RESULT} carries the rule identifier it answers, one of five status
codes, namely \texttt{OK}, \texttt{REJECTED}, \texttt{NOT\_AUTHORIZED}, \texttt{INVALID\_RULE},
and \texttt{NOT\_FOUND}, and an optional reason string.
A \texttt{NOT\_AUTHORIZED} response deliberately carries no reason, so a subscriber cannot probe the relay's authorization policy through error responses.

\subsection{Recommended Relay Resource Limits}
\label{app:limits}

\Cref{tab:resource_limits} gives concrete values for the per-session caps that \cref{subsec:rules} recommends.
The caps exist because a subscriber that installs rules is an authenticated participant at the transport level yet a potential adversary at the scheduling level, so the relay bounds what one session can consume before rule state or evaluation cost becomes an attack vector.
A multi-tenant relay additionally normalizes priorities across sessions, so one session's priority escalation cannot starve another's traffic.

\begin{table}[t]
\centering
\caption{Recommended per-session resource limits for a \momq relay.}
\label{tab:resource_limits}
\small
\begin{tabularx}{\columnwidth}{@{}l l L@{}}
\toprule
\textbf{Resource} & \textbf{Limit} & \textbf{Rationale} \\
\midrule
Rules per session & 100 & An order of magnitude above the four-rule policy of \cref{subsec:svc_rules} \\
Match entries per rule & 10 & Bounds the metadata one match decision inspects \\
Actions per rule & 20 & Bounds the merge work per directive \\
Key length & 128~bytes & Keys are short identifiers \\
Value length & 1024~bytes & Values may carry long identifiers \\
Installation rate & 10 per second & Adapts rules to path changes, defeats flooding \\
\bottomrule
\end{tabularx}
\end{table}

\subsection{Reconfiguration Event Duration}
\label{app:reconf_duration}

\Cref{fig:reconf_duration} gives the full duration distribution behind the bounded-outage claim of \cref{subsec:reconf_char}.
Across 103~events, durations range from 22~ms to 172~ms with a median of 58~ms and an interquartile range of 49--81~ms.
A 100~ms deferral window covers 88\% of events, so anticipating a reconfiguration and briefly steering around it costs little.

\begin{figure}[!t]
\centering
\includegraphics[width=0.85\columnwidth]{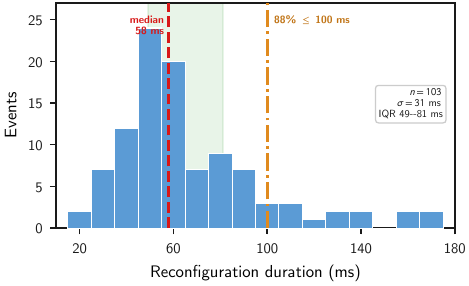}
\caption{Starlink reconfiguration event durations across 103~events in 30~minutes of 1~ms-rate probing.}
\label{fig:reconf_duration}
\end{figure}

\subsection{Single-Path Congestion-Control Baselines}
\label{app:cca_baselines}

\Cref{fig:cca_baselines} compares the congestion controllers of \cref{subsec:eval_transport} on each link in isolation, so the single-path baselines of the headline comparison rest on the best controller for each link rather than an arbitrary one.
On WiFi, Copa needs the lowest median playback buffer at 338.5~ms against 381.1~ms for BBRv3 and 410.8~ms for Cubic, and its FCT stays close behind at a 25.2~ms median and a 235.0~ms P99.
On Starlink, LeoCC needs the lowest buffer at 367.6~ms against 397.3~ms for Copa, 452.9~ms for BBRv3, and 475.1~ms for Cubic, and it holds the shortest tail with a P99.9 FCT of 541.7~ms against 681.1~ms for the next best.
The delay-based and LEO-aware controllers each win on their own link, so Copa carries the single-path WiFi baseline and LeoCC the single-path Starlink baseline throughout \cref{sec:evaluation}.

\begin{figure}[t]
\centering
\includegraphics[width=\columnwidth]{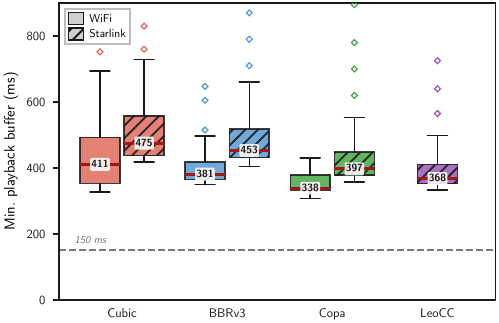}
\caption{Minimum playback buffer per congestion controller on each link in isolation, grouped by controller. Solid boxes are WiFi and hatched boxes are Starlink, and LeoCC runs on Starlink only. Boxes span the interquartile range with the median printed inside, whiskers the 1.5~IQR extent, and diamonds individual outliers. The dashed line marks the 150~ms interactive target, and the Starlink outliers reach 1.3~s.}
\label{fig:cca_baselines}
\end{figure}

\subsection{Rule Ablation Detail}
\label{app:ablation}

\subsubsection{Incremental Activation on the Live Testbed}
\label{app:ablation_live}

Incremental rule activation reveals complementary contributions (\cref{tab:live_ablation}).
The cost-sensitive default (\rfour) alone achieves a median buffer of 247.6~ms while using almost none of the backup path.
Adding \pframe interleaving (\rone) reduces the buffer to 198.4~ms by relieving head-of-line blocking during multi-round \iframe transmission.
Adding reconfiguration avoidance (\rtwo) provides the largest single improvement, with the buffer at 151.9~ms and P99.9 dropping from 233.7~ms to 117.5~ms, as collisions between \iframes and reconfigurations are proactively eliminated.
Dependency co-location (\rthree) brings the buffer to 127.9~ms by eliminating cross-path reordering of enhancement-layer frames.
Backup-path usage grows in step from 0.3\% to 11.3\%, because each rule sends the backup path exactly the traffic that benefits from it.

\begin{table}[t]
\centering
\caption{Incremental rule activation on the live testbed.}
\label{tab:live_ablation}
\small
\setlength{\tabcolsep}{3pt}
\begin{tabularx}{\columnwidth}{@{}L c c c@{}}
\toprule
\textbf{Rule set} & \textbf{Buffer} & \textbf{P99.9 FCT} & \textbf{Backup} \\
 & \textbf{(ms)} & \textbf{(ms)} & \textbf{use} \\
\midrule
Cost-sensitive default (\rfour) & 247.6 & 277.3 & 0.3\% \\
+ Interleaving (\rone) & 198.4 & 233.7 & 6.2\% \\
+ Reconf.\ avoidance (\rtwo) & 151.9 & 117.5 & 9.4\% \\
+ Dependency co-location (\rthree) & 127.9 & 114.1 & 11.3\% \\
\bottomrule
\end{tabularx}
\end{table}

\subsubsection{Controlled Simulation}
\label{app:ablation_sim}
\label{subsec:eval_simulation}

The live testbed cannot isolate mechanisms under reproducible conditions, so we complement it with a controlled simulation that isolates the steady-state contributions of \pframe interleaving (\rone) and dependency co-location (\rthree).
Reconfiguration events are not modeled, so reconfiguration avoidance (\rtwo) is evaluated only on the live testbed, and the cost-sensitive default (\rfour) remains fixed across all simulated configurations.

The simulation models two asymmetric paths whose parameters are drawn from the testbed characterization in \cref{sec:motivation}.
The primary path has a base delay of 40~ms, delay jitter of 5~ms, packet loss of 0.2\%, and capacity of 80~Mbps.
The backup path has a base delay of 15~ms, delay jitter of 2~ms, packet loss of 0.1\%, and capacity of 50~Mbps.
SVC parameters match the live setup, with LeoCC on the satellite path and Cubic on the backup.
Each of the 50~iterations executes five repetitions of four configurations in interleaved round-robin order, generating 250~runs per configuration.
The four configurations form a $2 \times 2$ factorial design around \rone and \rthree.

The primary metric is the \emph{p1--p99 jitter buffer}, the difference between the 99th and 1st percentiles of per-frame delivery deviation within each run, corresponding to the playback buffer needed to absorb 98\% of frame-level timing variation.
This metric is more robust to single-frame outliers than the minimum-maximum buffer used in the live evaluation, making it better suited to cross-run statistical comparison at scale.

\begin{table}[t]
\centering
\caption{Simulated playback buffer under rule removal ($n = 250$ runs per configuration).}
\label{tab:sim_ablation}
\small
\setlength{\tabcolsep}{3pt}
\begin{tabularx}{\columnwidth}{@{}L c c c c@{}}
\toprule
\textbf{Configuration} & \textbf{Median} & \textbf{$\sigma$} & \textbf{$\Delta$\%} & \textbf{Cohen's $d$} \\
 & \textbf{(ms)} & \textbf{(ms)} & & \\
\midrule
\momq baseline & 237.4 & 37.5 & n/a & n/a \\
w/o \rthree (co-location) & 342.2 & 80.9 & +44.3 & 1.77 \\
w/o \rone (interleaving) & 395.9 & 72.6 & +66.7 & 2.97 \\
w/o \rone and \rthree & 501.5 & 116.3 & +111.6 & 3.09 \\
\bottomrule
\end{tabularx}
\end{table}

With the baseline rule set, \momq achieves a median p1--p99 buffer of 237.4~ms (\cref{tab:sim_ablation}).
Removing dependency co-location increases the buffer by 44.3\%, as enhancement-layer frames split across paths with asymmetric RTTs force the receiver to buffer early-arriving frames.
Removing \pframe interleaving is more damaging at 66.7\%, which confirms that the convoy effect from multi-round \iframe transmission dominates tail latency.
Removing both yields a 111.6\% increase, almost exactly the sum of the individual degradations, since $44.3 + 66.7 = 111.0$.
This near-additive behavior confirms that interleaving and co-location address independent problems with negligible interaction.

\begin{figure}[!t]
\centering
\includegraphics[width=0.85\columnwidth]{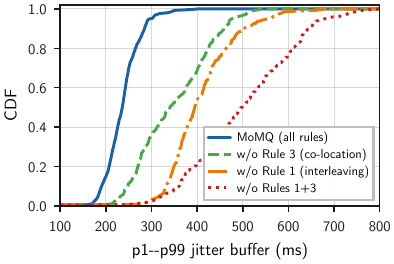}
\caption{CDF of p1--p99 jitter buffer across 250~runs per configuration.}
\label{fig:ablation_cdf}
\end{figure}

The CDF (\cref{fig:ablation_cdf}) shows four clearly separated distributions.
The 95th percentile of the baseline at 303~ms falls below the 5th percentile of the configuration without \rone and \rthree at 327~ms, so the two distributions overlap by less than 5\%.
Cohen's $d$ confirms large effect sizes above 1.7 for all baseline comparisons, and pairwise win rates reinforce the ordering.
\momq outperforms the variant without co-location in 90.0\% of run-level comparisons, the variant without interleaving in 99.1\%, and the variant without both in 99.3\%.

\begin{figure}[t]
\centering
\includegraphics[width=0.85\columnwidth]{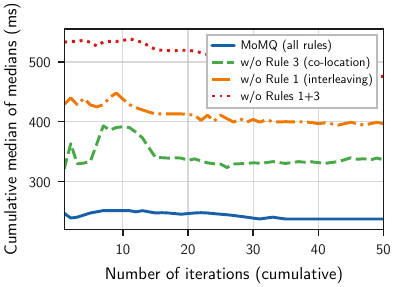}
\caption{Convergence of the cumulative median-of-medians over 50~iterations.}
\label{fig:ablation_convergence}
\end{figure}

\Cref{fig:ablation_convergence} tracks the cumulative median-of-medians across iterations, where every estimate converges to within $\pm$5~ms of its final value by roughly iteration~30 and remains stable through iteration~50.
The four curves maintain consistent separation throughout, ruling out the possibility that the ordering is an artifact of finite sample size.
Across 250~independent trials per configuration, interleaving contributes more than co-location, their effects are near-additive, and all pairwise comparisons are significant at $p < 0.001$ with Cohen's $d \geq 1.77$, so the rules address genuinely independent problems.

\begin{figure}[!t]
\centering
\includegraphics[width=\columnwidth]{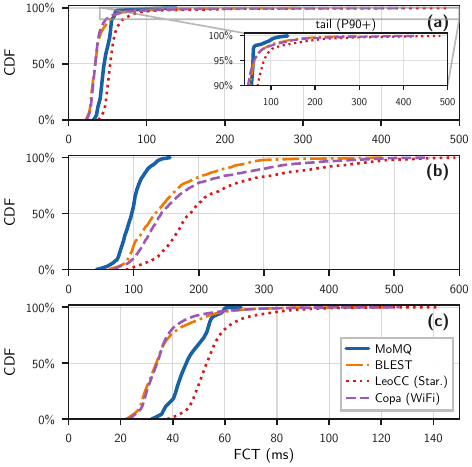}
\caption{FCT distributions with the Finland relay. (a)~All frames, with the tail above P90 magnified in the inset. (b)~\iframe FCT. (c)~\pframe FCT.}
\label{fig:finland_cdf}
\end{figure}

\begin{table}[!htb]
\centering
\caption{Hop alignment of the Starlink and terrestrial traceroutes from the Canada subscriber. Each row is one router with the hop at which each trace reaches it, and paired names give the Starlink and terrestrial unit where the traces hit adjacent equipment.}
\label{tab:canada_traceroute}
\small
\begin{tabularx}{\columnwidth}{@{}L c c l@{}}
\toprule
\textbf{Router} & \multicolumn{2}{c}{\textbf{Hop}} & \textbf{Match} \\
\cmidrule(lr){2-3}
 & \textbf{Starlink} & \textbf{Terr.} & \\
\midrule
\texttt{sea-b4-link} & & 12 & terrestrial only \\
\texttt{sea-b1-link} & 6 & 13 & interfaces differ \\
\midrule
\texttt{chi-bb2-link} & 7 & 14 & identical \\
\texttt{ewr-bb2} & 8 & 15 & identical \\
\texttt{ldn-bb2} & 9 & 16 & identical \\
\texttt{hbg-bb2} & 10 & 17 & identical \\
\texttt{hbg-b2} & 11 & 18 & identical \\
\midrule
\texttt{ic-383976} / \texttt{-75} & 12 & 19 & adjacent \\
\texttt{core22} / \texttt{core21} & 13 & 20 & adjacent \\
\texttt{core11} & 14 & 21 & identical \\
core spine & 15 & 22 & identical \\
\bottomrule
\end{tabularx}
\end{table}

\subsubsection{Finland FCT Distributions}
\label{app:finland_cdf}

\Cref{fig:finland_cdf} shows the full FCT distributions with the Finland relay, where the structural patterns of \cref{fig:fct_cdf} hold at $2.5\times$ the relay distance.
In \cref{fig:finland_cdf}a and its tail inset, the four distributions rise together through the body and separate in the tail, where the \momq curve terminates before 200~ms while the three baselines stretch toward 550~ms.
\Cref{fig:finland_cdf}b locates the separation in the \iframes, whose \momq curve sits left of every baseline across the whole distribution.
\Cref{fig:finland_cdf}c shows the price, since Copa and BLEST deliver the median \pframe slightly earlier than \momq, consistent with the medians of \cref{tab:finland_comparison}, yet the \momq \pframe distribution closes by roughly 70~ms while the baselines run past 100~ms.

\subsection{Traceroute Alignment for the Canada Experiment}
\label{app:traceroute}

\Cref{tab:canada_traceroute} aligns the raw hops of the two traceroutes behind the path-convergence analysis of \cref{subsec:eval_diversity}, with hostnames shortened to their router identifiers.
In the Seattle rows the traces reach the same router name on different interface addresses, 213.155.141.32 on the Starlink trace against 62.115.132.157 on the terrestrial one, and only the terrestrial trace passes \texttt{sea-b4-link}.
From \texttt{chi-bb2-link} at 62.115.132.154 onward the hostnames and addresses are identical on both traces, and the \texttt{bb2} hops expose the same MPLS label (415780) on both.
The datacenter ingress rows differ only in adjacent equipment before the traces share \texttt{core11} and the destination-side spine router.
Of the nine hops the Starlink trace takes from Chicago to the destination, seven reach the same router at the same address as the terrestrial trace and the other two reach adjacent equipment.

\end{document}